\documentclass[sigconf, screen]{acmart}

\usepackage{microtype}
\usepackage{booktabs} 
\usepackage{array}
\usepackage{multirow}
\usepackage{makecell}
\usepackage{colortbl}
\usepackage{textcomp}
\usepackage{paralist} 
\usepackage{cleveref}
\usepackage{graphicx}
\usepackage[tikz]{mdframed} 
\usepackage{url}
\usepackage{upgreek}

\usepackage[frozencache,cachedir=.]{minted}
\usepackage{siunitx}

\newcommand{\rot}[1]{\rotatebox{90}{\fontsize{8pt}{10pt}\selectfont{#1}}}

\renewcommand\acksname{Acknowledgements}

\newcommand\researchquestion[2]{
    \vspace{0.6em}
    \begin{mdframed}[
    	linewidth=1pt,
    	leftmargin=0pt,
    	topline=false,
    	rightline=false,
    	bottomline=false,
    	linecolor=gray!50]
    \begin{tabular}{@{}l m{0.89\textwidth}}
    	\textbf{{#1}} & {#2} \\
    \end{tabular}
    \end{mdframed}
    \vspace{0.3em}
}

\newcommand\RQAnswer[2]{
	\noindent 
	\begin{mdframed}[
		linewidth=1pt,
		leftmargin=0pt,
		backgroundcolor=gray!10,
		linecolor=gray!10]
		\begin{tabular}{@{}l m{0.89\textwidth}}
			\textbf{{#1}} & {#2} \\
		\end{tabular}
	\end{mdframed}
}

\definecolor{Conventional}{RGB}{31,120,180}
\definecolor{SelfEstimated}{RGB}{51,160,44}
\definecolor{Content}{RGB}{255,127,0}
\definecolor{Time}{RGB}{106,61,154}

\definecolor{CorrNone}{HTML}{FFFFFF}
\definecolor{CorrWeak}{HTML}{f0f0f0}
\definecolor{CorrMed}{HTML}{bdbdbd}
\definecolor{CorrStr}{HTML}{969696}

\definecolor{HighEfficacy}{HTML}{fcffa4}
\definecolor{LowEfficacy}{HTML}{4d0d6c}

\setcopyright{rightsretained}
\acmPrice{}
\acmDOI{10.1145/3540250.3549084}
\acmYear{2022}
\copyrightyear{2022}
\acmSubmissionID{fse22main-p44-p}
\acmISBN{978-1-4503-9413-0/22/11}
\acmConference[ESEC/FSE '22]{Proceedings of the 30th ACM Joint European Software Engineering Conference and Symposium on the Foundations of Software Engineering}{November 14--18, 2022}{Singapore, Singapore}
\acmBooktitle{Proceedings of the 30th ACM Joint European Software Engineering Conference and Symposium on the Foundations of Software Engineering (ESEC/FSE '22), November 14--18, 2022, Singapore, Singapore}

\begin{document}

\title[Correlates of Programmer Efficacy and Their Link to Experience]{Correlates of Programmer Efficacy and Their Link to Experience: A Combined EEG and Eye-Tracking Study}


\author{Norman Peitek}
\affiliation{%
  \institution{Saarland University,\\Saarland Informatics Campus}
  \city{Saarbr{\"u}cken}
  \country{Germany} 
}

\author{Annabelle Bergum}
\affiliation{%
  \institution{Saarland University,\\Saarland Informatics Campus,\\Graduate School of Computer Science}
  \city{Saarbr{\"u}cken}
  \country{Germany}
}

\author{Maurice Rekrut}
\affiliation{%
  \institution{German Research Center for Artificial Intelligence, Saarland Informatics Campus, Saarbr{\"u}cken, Germany}
  \city{}
  \country{}
}

\author{Jonas Mucke}
\affiliation{%
  \institution{Chemnitz University of Technology}
  \city{Chemnitz} 
  \country{Germany}
}

\author{Matthias Nadig}
\affiliation{%
  \institution{German Research Center for Artificial Intelligence, Saarland Informatics Campus, Saarbr{\"u}cken, Germany}
  \city{}
  \country{}
}

\author{Chris Parnin}
\affiliation{%
  \institution{NC State University}
  \city{Raleigh} 
  \state{North Carolina} 
  \country{USA}
}

\author{Janet Siegmund}
\affiliation{%
  \institution{Chemnitz University of Technology}
  \city{Chemnitz} 
  \country{Germany}
}

\author{Sven Apel}
\affiliation{%
  \institution{Saarland University,\\Saarland Informatics Campus}
  \city{Saarbr{\"u}cken}
  \country{Germany}
}

\begin{abstract}
\textbf{Background:} Despite similar education and background, programmers can exhibit vast differences in efficacy. While research has identified some potential factors, such as programming experience and domain knowledge, the effect of these factors on programmers' efficacy is not well understood.

\noindent\textbf{Aims:} We aim at unraveling the relationship between efficacy (speed and correctness) and measures of programming experience. We further investigate the correlates of programmer efficacy in terms of reading behavior and cognitive load. 

\noindent\textbf{Method:} For this purpose, we conducted a controlled experiment with 37~participants using electroencephalography (EEG) and eye tracking. We asked participants to comprehend up to 32~Java source-code snippets and observed their eye gaze and neural correlates of cognitive load. We analyzed the correlation of participants' efficacy with popular programming experience measures.

\noindent\textbf{Results:} We found that programmers with high efficacy read source code more targeted and with lower cognitive load. Commonly used experience levels do not predict programmer efficacy well, but self-estimation and indicators of learning eagerness are fairly accurate.

\noindent\textbf{Implications:} The identified correlates of programmer efficacy can be used for future research and practice (e.g., hiring). Future research should also consider efficacy as a group sampling method, rather than using simple experience measures.
\end{abstract}



\maketitle
\renewcommand{\shortauthors}{N. Peitek, A. Bergum, M. Rekrut, J. Mucke, M. Nadig, C. Parnin, J. Siegmund, S. Apel}

\section{Introduction}

Proficient programmers are essential for providing the critical infrastructure and functioning applications for our modern society~\cite{Garousi2019}. Although the learning strategies, education, and pathways to becoming a programmer may differ, the general expectation is that, with more time and experience, a programmer should generally become more proficient. For example, many research studies of programmers use years of experience~\cite{Latoza2007, Bednarik2012}, education level~\cite{Crosby2002, Peachock2017, Jbara2017}, or employment status~\cite{Lee2016, Busjahn2015, Burkhardt2002} as foundational measures for proficiency. While all these choices are sensible, they all implicitly encode the expectation that higher proficiency should correlate with more experience.

However, in practice, reported observations violate this expectation. Hiring managers and technical founders~\cite{atwood2022} report that they routinely encounter ``engineers with years of experience who couldn't competently program''~\cite{hackerNews}, who ``struggle with tiny problems''~\cite{ghory2022}, and ``\emph{Senior Engineers} who can't write basic code''~\cite{myth2020}. The original creator of the infamous \emph{FizzBuzz} interview question~\cite{ghory2022} only did so after seeing that the ``majority of comp sci graduates... and self-proclaimed senior programmers'' had difficulty solving simple problems in a timely manner. Research has also found that, in companies, the seniority level showed little correlation to actual programming skill~\cite{Jorgensen2020}, and programmers with similar education and background can exhibit vast differences in productivity, up to factors of ten~\cite{McConnell2011}.
Research is still at a loss when it comes to explaining the cause of these differences, accounting for different
trajectories of learning that underlie programming education and training~\cite{Exter2018, Groeneveld2021}, or identifying proficient programmers during a hiring process~\cite{Li2015}. 

In this paper, we explore the idea of unraveling programmer \emph{efficacy}\footnote{Efficacy specifically refers to the ability to quickly produce the intended result. Notably, this differs from expertise, which involves additional facets, such as deep knowledge, effort, and mastery of skills.}---based on speed and correctness---with the help of programmer \emph{experience}---amount of learning or practice. To improve our understanding of the relationship between efficacy and experience, we conducted a combined electroencephalography (EEG) and eye-tracking study, allowing us to take a close look at how differences in efficacy and experience are related to cognitive differences among programmers. In our study, 37~participants with varying levels of experience performed program-comprehension tasks. We found that programmers with higher efficacy read code more targeted, with shorter fixations, fewer (re)fixations, and skipping more code elements. They also complete their tasks with lower cognitive load, in less time, and make fewer errors than programmers with lower efficacy. Interestingly, we found that commonly used experience measures do not correlate with the observed efficacy, but instead self-estimation and learning indicators have considerable predictive power. To this end, we have identified correlates of high programmer efficacy as well as experience measures that provide a strong link to efficacy.

In summary, we make the following contributions:
\begin{compactitem}
\item A combined EEG and eye-tracking experiment to investigate programmer efficacy with a diverse participant pool.
\item Confirmation of prior results that programmers with high efficacy read source code more efficiently and with lower cognitive load.
\item Empirical evidence that conventional experience measures have only poor predictive power for programmer efficacy. Self-estimation and indicators of learning eagerness are better suited.
\item An online replication package\footnote{\url{https://github.com/brains-on-code/NoviceVsExpert}}, including experiment design, raw data, and executable analysis scripts.
\end{compactitem}

\section{Research Questions and Variables}
\label{sec:ResearchQuestions}

\begin{table*}[ht]
\caption{Overview of our research questions and chosen measures.}
\label{tab:ResearchQuestions}

\fontsize{9pt}{9pt}\selectfont
\begin{tabular}{lll}
\toprule
RQ & Literature & Selected Measure(s) \\ \midrule
\multirow{2}{*}{RQ\textsubscript{1}} & Al Madi et al.~\cite{AlMadi2021} & \makecell[l]{Fixation durations (singe fixation duration, total fixation duration) and fixation pro-\\babilities (skipping probability, single fixation probability, multiple fixation probability)}\\
 & Aljehane et al.~\cite{Aljehane2021}, Busjahn et al.~\cite{Busjahn2015} & Code element coverage, code element coverage over time\\
\specialrule{0.1pt}{1pt}{1pt}
\multirow{2}{*}{RQ\textsubscript{2}} & Lee et al.~\cite{Lee2016} & Alpha, beta, theta, and gamma power \\
& Medeiros et al.~\cite{Medeiros2021}, Holm et al.~\cite{Holm2009} & Ratio of theta power at frontal brain region and alpha power at parietal brain region  \\
\specialrule{0.1pt}{1pt}{1pt}
RQ\textsubscript{3} & & \makecell[l]{Correlation between efficacy and experience measures} \\ 
\bottomrule
\end{tabular}
\end{table*}

Our study on programmer efficacy builds on the methodology of previous experiments investigating program comprehension and programming experience. Our aim is to incorporate measures of program comprehension from these experiments into a single coherent study as shown in~\Cref{tab:ResearchQuestions}. We specifically designed our study to better understand programmer efficacy across a wide range of experience levels in the context of program-comprehension tasks. Programmer efficacy is therefore the independent variable for our experiment. We operationalize programmer efficacy as follows:

\vspace{1em}
$\mathit{programmer~efficacy} = \frac{\mathit{number~of~correct~code~comprehension~tasks}}{\mathit{completion~time~in~minutes}}$\\

Note that programmer efficacy captures both the speed and correctness of a participant's behavior. Due to the strict one-hour time limit of the experiment~(see~\Cref{sec:ExperimentPlan}), only fast participants were able to see all snippets before the experiment ended. Our definition of programmer efficacy eliminates effects arising from a difference in the number of attempted tasks and is in line with prior work~\cite{Shneiderman1979, Wiedenbeck1985, Hofmeister2019}.

To guide our experiment design regarding dependent variables, we defined several research questions, which we introduce next.

\subsection{Reading Behavior (RQ\textsubscript{1})}
\label{sec:ResearchQuestionsRQ1}

Several studies have shown that experienced programmers show a different reading behavior than novices based on eye-gaze measures. The reading behavior describes a programmer's eye movements while they are comprehending a source-code snippet. Therefore, we state the following research question:

\researchquestion{RQ\textsubscript{1}}{Do different levels of programmer efficacy exhibit differences in reading behavior (in terms of navigation strategy and code element coverage)?}

\paragraph{Operationalization}

For RQ\textsubscript{1}, we adopt 7 measures that have been identified in the literature: In a longitudinal study, Al Madi et al. observed differences in the navigation strategy at the token level~\cite{AlMadi2021}. Specifically, they analyzed fixation duration and how likely it was that a participant (re)fixated on a token. Experts showed significantly shorter fixation durations, a lower chance of refixating on a token, and a higher chance to skip tokens. Likewise, Aljehane et al. found differences between novices and experts in terms of \emph{code element coverage}~\cite{Aljehane2021}. It refers to the number of elements on which a participant fixates during a task in contrast to the total number of code elements. They found that novices read, in particular, more method signatures, variable declarations, identifiers, and keywords. Finally, Busjahn et al. have also identified a difference in code element coverage, in that novices fixate more code elements than experts~\cite{Busjahn2015}. They also identified a difference in the \emph{reading order}. However, our replication study revealed that this effect is principally driven by the execution order of the snippets~\cite{Peitek2020:Linearity}. As our snippets are not balanced for this aspect, we will not consider reading order as a measure; these studies used different measures to pinpoint the participants' reading behavior. In our study, we aim at answering how programmer efficacy affects reading behavior across all described measures to increase comparability.

\subsection{Cognitive Load (RQ\textsubscript{2})}

\vbox{
Similar to reading behavior, previous studies found that different levels of programming experience can be distinguished by the observed neural correlates of cognitive load, which leads us to our next research question:

\researchquestion{RQ\textsubscript{2}}{Do different levels of programmer efficacy exhibit differences in neural correlates of cognitive load?}
}

\paragraph{Operationalization}

Related to RQ\textsubscript{2}, Crk et al. used event-related desynchronization along alpha and theta powerbands measured by an EEG device to classify participants into two programming experience levels~\cite{Crk2014}. Similarly, Lee et al. found a similar effect also using EEG, but using a more comprehensive analysis across more EEG bands~\cite{Lee2016}. Ishida and Uwano found an increase in the alpha frequency band for programmers who successfully finished tasks~\cite{Ishida2019EEG}. However, Madeiros et al. compared several EEG measures to distinguish different experience levels and suggested the ratio between theta and alpha power as a measure of cognitive load~\cite{Medeiros2021}. Notably, this ratio is commonly used as a cognitive load measure in other fields~\cite{Holm2009, Kartali2019}. We analyze both powerbands and the ratio measure in our study to better understand the link between programmer efficacy and cognitive load.

\subsection{Experience Measures (RQ\textsubscript{3})}

Finally, we focus on programmer efficacy as a distinguishing factor between our participants. Prior research most commonly uses rather simple experience measures to separate participants (e.g., years of programming). In this vein, we pose our last research question:

\researchquestion{RQ\textsubscript{3}}{Do different levels of programmer efficacy correlate with common measures of programming experience?}

\paragraph{Operationalization}

Many commonly used experience measures can be inconsistent in their predictive power, which led us to use programmer efficacy as the distinguishing factor. In RQ\textsubscript{3}, we analyze widely used experience measures\footnote{We share the full list of experience measures on the project's Web site.} and distill the relevant measures that correlate best with the observed programmer efficacy.

\section{Study Design and Conduct}

To answer our research questions, we designed and conducted a study, which we describe in this section. All materials including snippets, tasks, and the experiment script, are available on the project's Web site.$^2$

\subsection{Experiment Plan}
\label{sec:ExperimentPlan}

\begin{figure*}[ht]
    \centering
    \includegraphics[trim={0 14cm 13.5cm 0}, clip, width =0.9\textwidth]{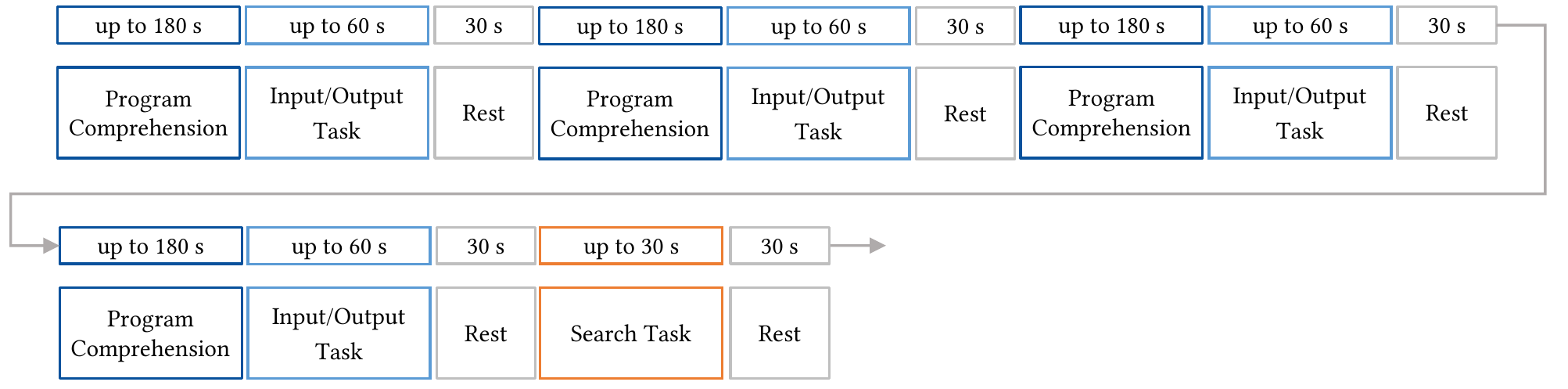}
    \caption{Visualization of the experiment design, which shows the first 4 comprehension tasks. The experiment ended after all 32~comprehension tasks were completed, or, at latest, after 60~minutes. Participants could take a voluntary break after 20~minutes and 40~minutes.}
    \label{fig:experimentdesign}
\end{figure*}

We opted for a within-subject experiment design~\cite{Charness2012Experimental}, which we illustrate in~\Cref{fig:experimentdesign}. We presented up to 32 source-code snippets with a program-comprehension task (cf.~\Cref{sec:ComprehensionTask}), which were pseudo-randomized to avoid learning and fatigue effects. The experiment ended after the 32 program-comprehension tasks were completed or after 60~minutes, whichever happened first. The tasks were presented in three runs of 20~minutes and a voluntary break of 5 minutes between runs. In addition to the program-comprehension tasks, we included a search task, in which participants had to count brackets. This serves as a baseline for neural activation and is common in neuroscience studies of program comprehension~\cite{Siegmund2017, Peitek2021:Complexity}. We chose a 4:1 design, so a participant completes four program-comprehension tasks and one search task. Between tasks, we included a 30~second rest condition, in which participants were instructed to focus their eyes on a fixation cross and relax.

For the comprehension tasks, the participants could choose among four answer options (cf.~\Cref{ls:code_snippet}) as well as the option ``next'' to skip a task. Participants used the left hand to press space as ``submit'' and ``continue'' buttons, and the right hand to navigate with the arrow keys between answer options. After a short training session on the experiment flow (i.e., presentation of an example snippet, example input/output task, answering possible clarification questions), participants could use the keyboard without constantly looking at their hands, which minimizes motion artifacts.

Our study was approved by the ethical review board of the faculty of Mathematics and Computer Science at Saarland University.

\subsection{Snippet Selection}
\label{sec:SnippetSelection}

\begin{table}[]
\caption{Overview of the source-code snippets used in the experiment and the outcomes in terms of correctness and response time. The number of times a snippet was shown is unbalanced due to the randomized presentation and outlier removal~(cf.~\Cref{sec:OutlierRemoval}).}
\label{tab:SnippetOverview}

\fontsize{9pt}{9pt}\selectfont
\begin{tabular}{lrr}
\toprule
Snippet & Correctness & Response Time \\
& (in \%) & (in sec, Mean $\pm$ SD) \\
\midrule
Ackermann & 5/36\,\,\,(14\%) & 83.51 $\pm$ 53.16 \\
ArrayAverage & 30/35\,\,\,(86\%) & 31.29 $\pm$ 17.03 \\
BinarySearch & 13/35\,\,\,(37\%) & 55.57 $\pm$ 31.16 \\
BinomialCoefficient & 10/32\,\,\,(31\%) & 77.70 $\pm$ 39.83 \\
BinaryToDecimal & 13/35\,\,\,(37\%) & 62.18 $\pm$ 25.27 \\
BogoSort & 24/35\,\,\,(69\%) & 79.64 $\pm$ 42.77 \\
CheckIfLettersOnly & 32/33\,\,\,(97\%) & 36.81 $\pm$ 22.69 \\
ContainsSubstring & 31/32\,\,\,(97\%) & 56.36 $\pm$ 29.74 \\
DropNumber & 15/36\,\,\,(42\%) & 55.19 $\pm$ 25.67 \\
GreatestCommonDivisor & 29/31\,\,\,(94\%) & 70.86 $\pm$ 32.40 \\
HeightOfTree & 26/34\,\,\,(76\%) & 41.81 $\pm$ 23.63 \\
hIndex & 19/33\,\,\,(58\%) & 100.60 $\pm$ 44.14 \\
InsertionSort & 34/35\,\,\,(97\%) & 67.40 $\pm$ 38.21 \\
IsAnagram & 26/30\,\,\,(87\%) & 101.70 $\pm$ 44.67 \\
IsPrime & 30/35\,\,\,(86\%) & 21.67 $\pm$ 12.91 \\
LengthOfLastWord & 33/36\,\,\,(92\%) & 64.52 $\pm$ 21.13 \\
MedianOnSorted & 23/30\,\,\,(77\%) & 45.44 $\pm$ 22.32 \\
Palindrome & 35/35\,\,(100\%) & 26.26 $\pm$ 14.57 \\
PermuteString & 21/35\,\,\,(60\%) & 129.24 $\pm$ 50.06 \\
Power & 29/33\,\,\,(88\%) & 34.08 $\pm$ 15.59 \\
RabbitTortoise & 11/33\,\,\,(33\%) & 103.54 $\pm$ 46.45 \\
RecursivePower & 29/31\,\,\,(94\%) & 23.68 $\pm$\,\,\,\,9.18 \\
Rectangle & 21/29\,\,\,(72\%) & 26.28 $\pm$ 18.81 \\
RemoveDoubleChar & 30/33\,\,\,(91\%) & 40.98 $\pm$ 17.68 \\
ReverseArray & 30/33\,\,\,(91\%) & 40.73 $\pm$ 23.21 \\
ReverseQueue & 26/33\,\,\,(79\%) & 41.26 $\pm$ 26.64 \\
SieveOfEratosthenes & 19/33\,\,\,(58\%) & 105.38 $\pm$ 48.16 \\
SignChecker & 31/33\,\,\,(94\%) & 28.53 $\pm$ 12.04 \\
SmallGauss & 18/36\,\,\,(50\%) & 27.24 $\pm$ 16.68 \\
SumArray & 32/33\,\,\,(97\%) & 13.70 $\pm$\,\,\,\,7.58 \\
UnrolledSort & 30/34\,\,\,(88\%) & 61.85 $\pm$ 32.32 \\
Vehicle & 34/35\,\,\,(97\%) & 36.13 $\pm$ 19.43 \\
\specialrule{0.1pt}{1pt}{1pt}
\specialrule{0.1pt}{1pt}{1pt}
Overall & 789/1072\,\,\,(74\%) & 56.02 $\pm$ 41.63\\
\bottomrule
\end{tabular}
\end{table}

\begin{listing}
\caption{Example source-code snippet with intermediate complexity that checks a string for the existence of a substring. An example input provided to the participant would be \texttt{containsSubstring("Example","Sample")} with answer options ``\texttt{Always False}'', ``\texttt{False}'' (correct), ``\texttt{True}'', and ``\texttt{3}''.}
\label{ls:code_snippet}
\begin{minted}[
    fontsize=\fontsize{7pt}{7pt},
    linenos,
    xleftmargin=10pt,
    numbersep=5pt,
    frame=lines]{Java}
public boolean containsSubstring(String word, String substring) {
    boolean containsSubstring = false;

    for (int i = 0; i < word.length(); i++) {
        for (int j = 0; j < substring.length(); j++) {
            if (i + j > word.length()) {
                break;
            }
            if (word.charAt(i + j) != substring.charAt(j)) {
                break;
            } else {
                if (j == substring.length() - 1) {
                    containsSubstring = true;
                    break;
                }
            }
        }
    }

    return containsSubstring;
}
\end{minted}
\end{listing}

\noindent
A crucial element of our experiment are the source-code snippets. We aimed at selecting snippets covering a variety of complexities. This ranges from simple snippets, with only a few lines that can be understood within seconds, to complex source-code snippets that require substantial mental effort to comprehend. Thus, our snippets require different levels of cognitive effort to comprehend them. This helps us to comprehensively capture program comprehension in relation to programmer efficacy.

We started our search by selecting Java snippets from a variety of previous studies of program comprehension~\cite{Siegmund2017, Peitek2021:Complexity, Busjahn2015}. We then searched for further snippets implementing algorithms of comparable complexity (in terms of size, nesting depth, and execution length), which lead to a pool of 38~snippets. Three of the authors independently assessed each snippet's complexity and suitability for the study (e.g., whether prior knowledge is necessary to comprehend it). While the snippets are in Java, we aimed at selecting snippets that did not require deep knowledge of the language. This resulted in 32~snippets. We show a sample snippet that checks for the existence of a substring in~\Cref{ls:code_snippet}. We list all included snippets in~\Cref{tab:SnippetOverview} and provide them in the replication package.

\subsection{Program-Comprehension Task}
\label{sec:ComprehensionTask}

We subdivided the program-comprehension task into two distinct steps to isolate the underlying cognitive processes of program comprehension and mental calculation of the result. To this end, we first presented the source-code snippet and instructed participants to comprehend it. Once a participant confirmed they comprehended its functionality (by pressing a button), we presented a sample input. Then, the participant had to mentally calculate the resulting output. The sample input was not shown until after the participant fully comprehended the code snippet, to prevent premature mental calculation without fully understanding the snippet. We informed participants of this multiple-step process beforehand and ensured their understanding with two training snippets.

\subsection{Experiment Execution and Data Collection}
\label{sec:ExperimentExecution}

All participants provided their informed consent and completed our experience questionnaire (see~\Cref{sec:Participants}).
We put the EEG cap on the participant's head, calibrated the eye-tracker, and started the experiment.
After the experiment, we conducted a semi-structured interview, including questions on their subjective views about the experiment as well as each snippet.

The EEG laboratory is in a dim-light room with minimized distractions, such as external sounds or mobile devices. Participants sat in a comfortable chair to prevent unnecessary muscle movements to reduce noise and artifacts in the EEG signal.
EEG signals were recorded using \emph{LiveAmp 64}\footnote{Brain Products GmbH, \url{https://brainvision.com/products/liveamp-64/}}, which is a 64-channel EEG device.
The sampling rate was set to 500~Hz, and the international 10--20 system of electrode placement~\cite{Jasper1958} was used to cover the entire scalp and obtain spatial information from the brain recordings.
For simultaneous eye tracking, we used the Tobii Pro Fusion eye-tracker\footnote{Tobii AB, \url{https://www.tobiipro.com/product-listing/fusion/}} attached to the screen. The eye gaze was recorded with a sampling rate of 250~Hz. The experiment was run with a PsychoPy~\cite{Peirce2019Psychopy} script (available in the replication package).

\subsection{Participants}
\label{sec:Participants}

\begin{table}[]
\caption{Overview of our participant demographics and 8 out of 49 overall measures from our experience questionnaire.\\
$^{\alpha}$ denotes self-estimated measures on a 1--5 Likert scale~\cite{Likert1932}.\\
$^{\beta}$ denotes measures that were only collected from (at least part-time) professional programmers.}
\label{tab:demographics}

\begin{tabular}{@{}lr@{}}
\toprule
Participant Demographic/Experience Measure & No./Mean $\pm$ SD \\
\midrule
Number of (Included) Participants & 37 \\
Female & 5 \\
Male & 31 \\
Non-Binary & 1 \\
\specialrule{0.1pt}{1pt}{1pt}
Age (in Years) & 25.95 $\pm$ 6.76 \\
\specialrule{0.1pt}{1pt}{1pt}
Undergraduate/graduate students & 27 of 37 \\
\dots~of which work (at least part time) & 14 of 27 \\
Full-time professionals & 10 of 37 \\
\specialrule{0.1pt}{1pt}{1pt}
Years of Learning Programming & 7.93 $\pm$ 6.14 \\
Years of Professional Programming & 3.55 $\pm$ 4.30 \\
Years of Java Programming & 4.54 $\pm$ 4.31 \\
Number of Known Programming Languages & 5.11 $\pm$ 2.02 \\
\specialrule{0.1pt}{1pt}{1pt}
Comparison to Peers$^{\alpha}$ & 3.67 $\pm$ 0.76\\
Comparison to 10-Year Professional$^{\alpha}$ & 2.25 $\pm$ 0.94\\
\specialrule{0.1pt}{1pt}{1pt}
Hours per Week Spent in Software Engineering$^{\beta}$ & 24.76 $\pm$ 21.08\\
Hours per Week Spent Programming$^{\beta}$ & 10.78 $\pm$ 11.36\\
\bottomrule
\end{tabular}
\end{table}

We recruited participants at Saarland University via e-mail lists and online bulletin boards. Participants received 10~Euro compensation each.
The prerequisite for participation was, at least, one year of experience with Java or, at least, three years of experience with a related programming language, such as \texttt{C\#}. It was important to recruit a diverse set of programmers with a wide range of experience levels to explore the differences and commonalities across different efficacy levels. In~\Cref{tab:demographics}, we provide an overview of our participants' demographics. Based on the conventionally used experience measures, our participant sample exhibits a wide range of programming experience, from programmers in their first year of programming at the university to 30~years of experience in industry.

We invited 39~participants to start the experiment. 38 out of the 39~measured participants completed the experiment.\footnote{One participant aborted the experiment early due to personal time constraints.} For one participant, the eye-tracker could not be calibrated, but all other modalities are available and the data are included in the analysis. We later excluded one complete participant data set due to their behavioral data consisting of a majority of outliers~(cf.~\Cref{sec:OutlierRemoval}), leaving us with 37~participants included in the data set. We based our programming experience questions on a validated questionnaire~\cite{Siegmund2014:Exp}, but extended it to cover more topics for further investigation~(e.g., programming-content consumption and production and daily work; the questionnaire is available in the replication package).

\section{Data Analysis}
\label{sec:dataanalysis}

In this section, we describe our data-analysis procedure for eye-tracking data, EEG data, and experience measures.

\subsection{Outlier Removal}
\label{sec:OutlierRemoval}

We started with removing outliers in response time for comprehending a source-code snippet. Specifically, we discarded the slowest 5\% (i.e., a response time of over 2~min~32~s) as well as the fastest 5\% (under 11~s), but only if a participant chose the option ``Next'', since some tasks can be rapidly answered by proficient programmers (e.g., the \texttt{SignChecker} snippet). This way, data points were only removed when participants did not thoroughly attempt to understand the snippet and program comprehension may not have occurred. 
With these rules, we removed 9~data points\footnote{One data point here is the response time for one participant to one snippet} in the lower 5\% and 55 in the upper 5\%, leaving 1072 data points for further analysis.
This led to the removal of more than half of the data points for one (slow) participant, so we excluded the participant's entire data set from the study.

\subsection{Eye-Tracking Data}

We defined areas-of-interests (AOIs) for each source-code snippet to relate eye-gaze behavior with particular regions and elements of code. For this purpose, we first obtained the abstract syntax tree of each source-code snippet to identify each token and code element. Then, we manually identified higher-order syntactical structures such as the head and body of loops or if-else-statements for each snippet and thereupon defined AOIs based on categories defined in previous work~\cite{Aljehane2021, AlMadi2021}.

We preprocessed the eye-tracking data to separate fixations and saccades from the raw stream of \texttt{(x,y)} coordinates. For this purpose, we used \texttt{I2CM}, which is a noise-robust algorithm~\cite{Hessels2017}. Then, we computed the token-level and element-coverage measures described in~\Cref{sec:ResearchQuestionsRQ1}.

\subsection{EEG Data}
\label{sec:EEGAnalysis}

Our cognitive load measures are based on a spectral analysis of EEG data. To improve the robustness of our analysis, we calculate cognitive load using two different approaches: theta/alpha power ratio and relative power analysis.

\subsubsection*{Data Cleaning}

First, we removed noise from the EEG data, especially movement artifacts, using established preprocessing methods: We filtered the data using Hamming-windowed finite impulse response (FIR) filters.
Power line noise was removed by a notch filter with a lower cutoff frequency of \SI{49}{\hertz} and an upper cutoff frequency of \SI{51}{\hertz}.
Second, to obtain a more robust performance in the subsequent analysis, we also removed baseline drifts and high-frequent noise with a bandpass filter ranging from 4 to \SI{200}{\hertz}~\cite{Rekrut2020Decoding}.

Due to the nature of the experiment, where participants have to look at different points of the screen, eye movements and other muscle activity cannot be fully avoided.
Therefore, we performed blind source separation as a second step to remove corresponding artifacts.
The used \texttt{EEGLAB} toolbox~\cite{Delorme2004}\footnote{Version 2021.1, \url{https://sccn.ucsd.edu/eeglab/}} provides an automated implementation of the independent component analysis (ICA).
For each component, it yields a classification with a confidence level, which we can use to automatically reject noisy components.
Rejection thresholds for eye and muscle artifacts were set to the default values of 0.9, while components without a clear assignment to a group were rejected at 0.95.

\subsubsection*{Theta/Alpha Power Ratio}

As a first technique, we computed the cognitive load based on the work of Holm et al.~\cite{Holm2009} and Kartali et al.~\cite{Kartali2019} as the ratio of the relative power of the theta and alpha band of the EEG. We calculated the power spectral density on a sliding window obtaining the mean power within each frequency band of interest.
The sliding window had a size of \SI{3}{\second} and was shifted by \SI{0.1}{\second} over each task. The cognitive load measure is obtained by dividing the mean power within a window. By using the sliding window, we can observe the time course of the cognitive load during the tasks.

\subsubsection*{Relative Power Analysis}

As a second technique, we also computed a per-band and per-channel relative power analysis along alpha, beta, gamma, and theta band based on the technique established by Lee et al.~\cite{Lee2016}, which is sensitive to the spatial location of brain activation by considering the channel of the electrode.\footnote{To facilitate comparison with Lee et al.'s work~\cite{Lee2016}, which used two groups, we formed two groups post-hoc based on experimental scores of efficacy. We separated our participants by performance into thirds. For increased potential, we contrasted the high-performing group (i.e., efficacy >= 0.35, n=12) and the low-performing group (i.e., efficacy <= 0.29, n=13), leaving out the middle group.}

\subsection{Efficacy and Experience}

For RQ\textsubscript{3}, we correlated the observed efficacy with all collected numerical and categorical experience measures. To this end, we used Spearman's~$\rho$ correlation due to the mix of continuous and ordinal nature of our data.

\section{Results and Discussion}
In this section, we present the results of our EEG and eye-tracking data analysis, along with their subsequent interpretation.

\begin{figure}[ht]
    \centering
    \includegraphics[trim={0 0 0.25cm 0}, clip, width =\columnwidth]{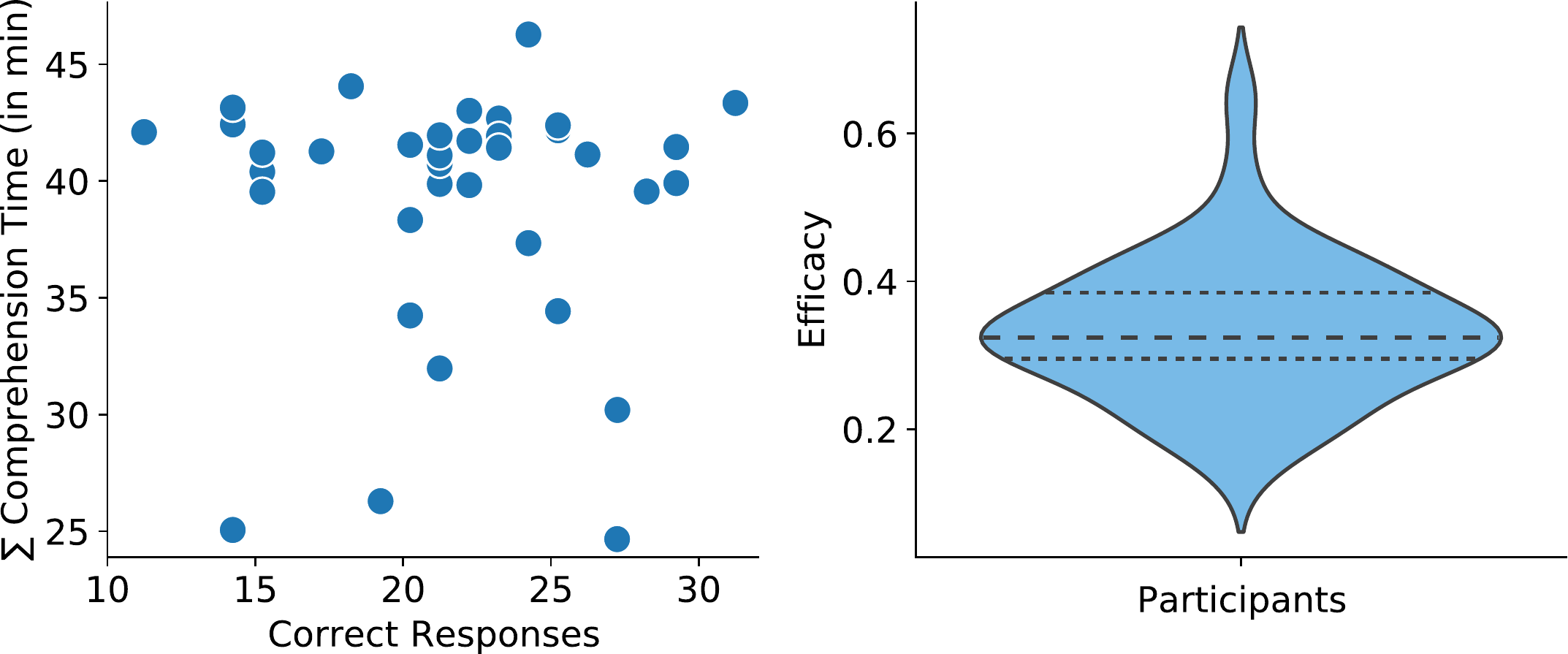}
    \caption{Illustration of behavioral data and the resulting distribution of programmer efficacy among our participants.}
    \label{fig:EfficacyDistribution}
\end{figure}

Applying our programmer efficacy definition to the data, we find that, on average, 1~task was correctly solved around every 3 minutes (i.e., mean programmer efficacy of 0.33 $\pm$ 0.10). We visualize the distribution of participant efficacy in~\Cref{fig:EfficacyDistribution}, which is a non-normal distribution according to a Shapiro-Wilk test ($W = 0.94, p = 0.046$). It could be normalized by removing the upper tail (2 participants)---neither which were the most experienced programmers---but since we already removed outliers, this appears to be the real distribution in our sample.

\subsection{Reading Behavior (RQ\textsubscript{1})}

\begin{figure*}[h]
\includegraphics[width=0.65\columnwidth]{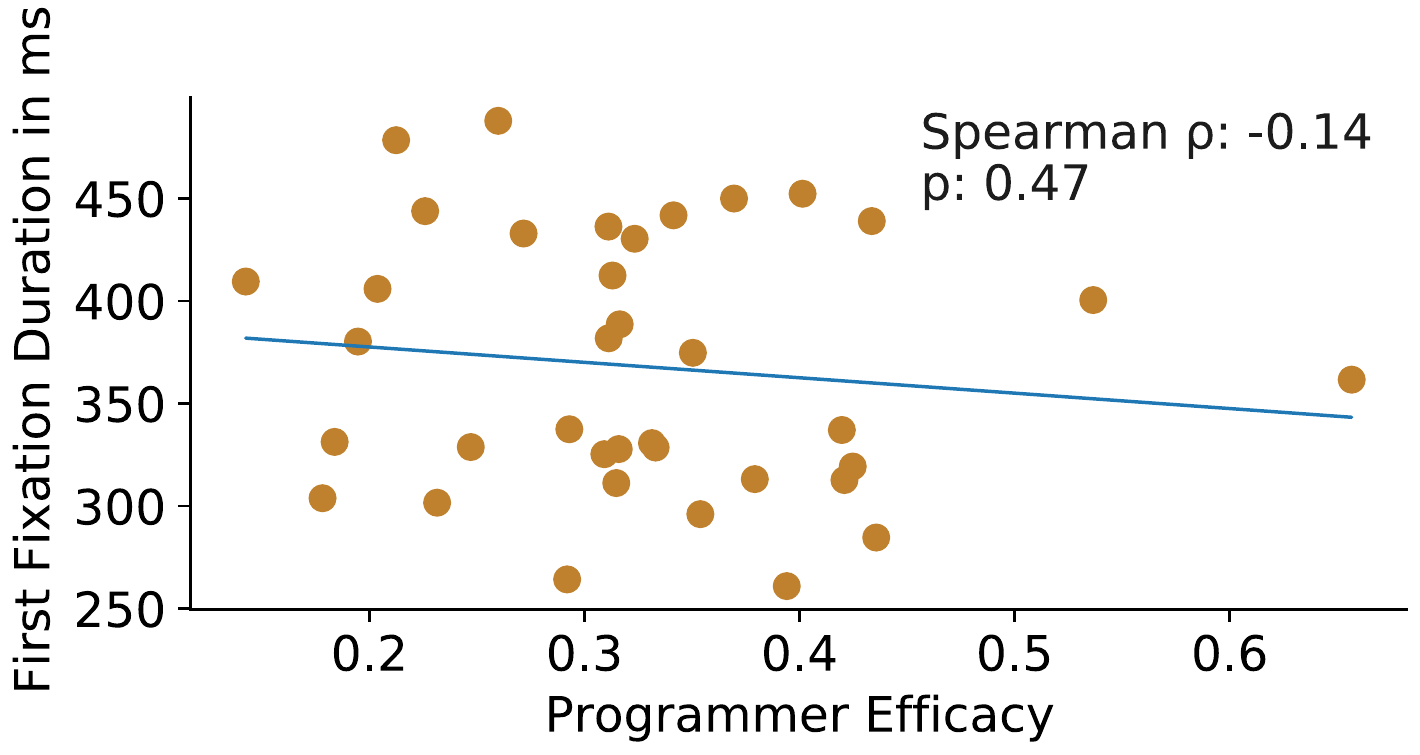}
\hspace{10pt}
\includegraphics[width=0.65\columnwidth]{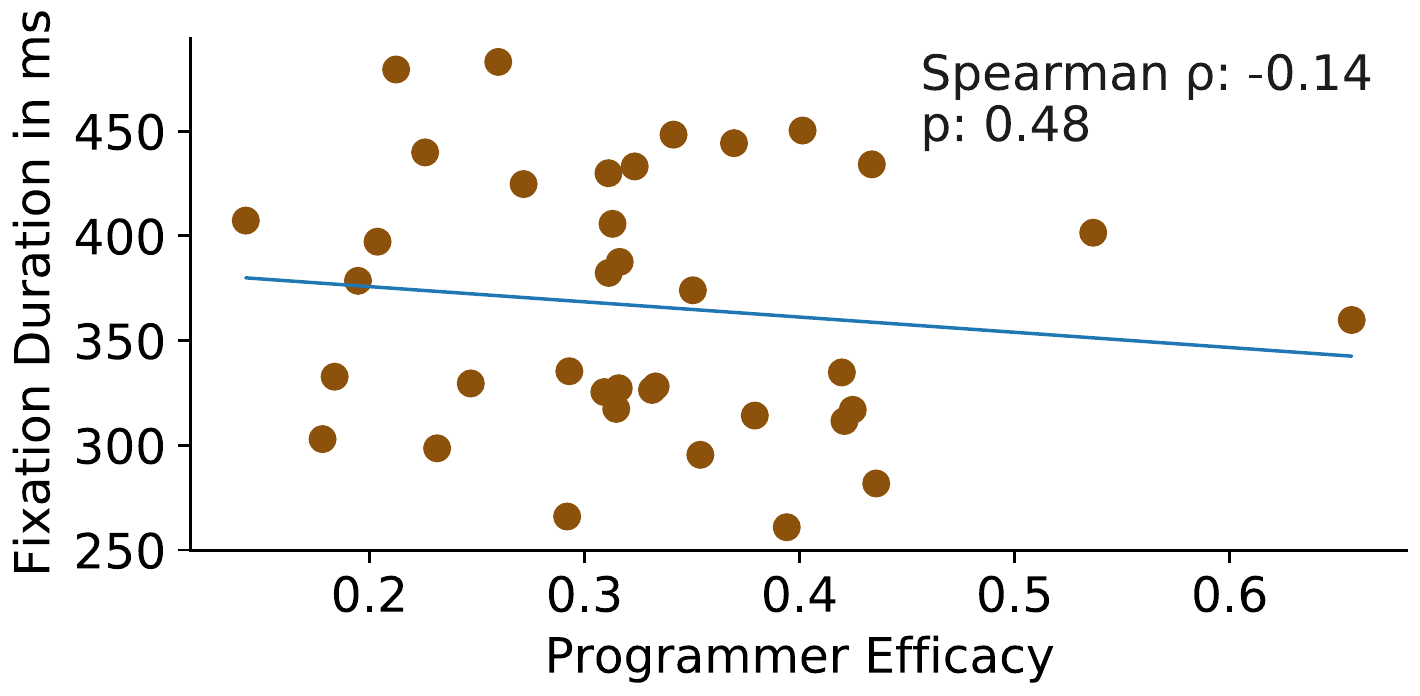}
\hspace{10pt}
\includegraphics[width=0.65\columnwidth]{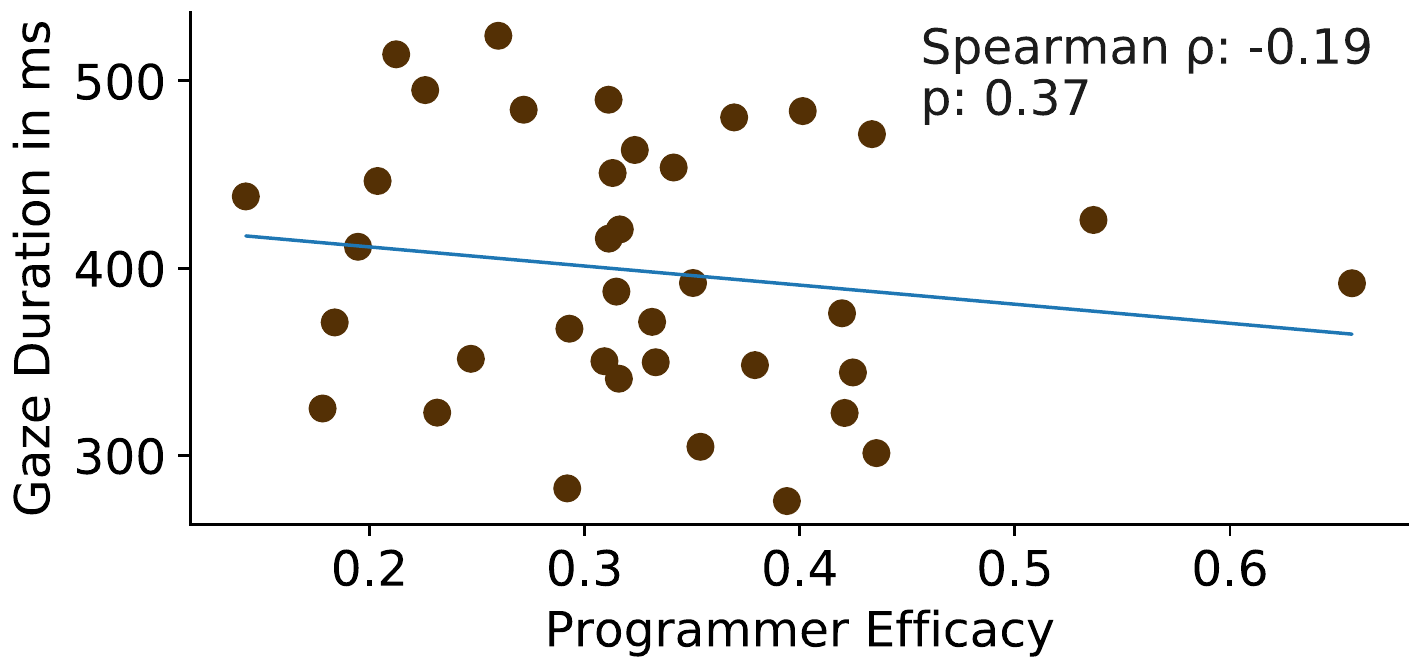}\\

\includegraphics[width=0.65\columnwidth]{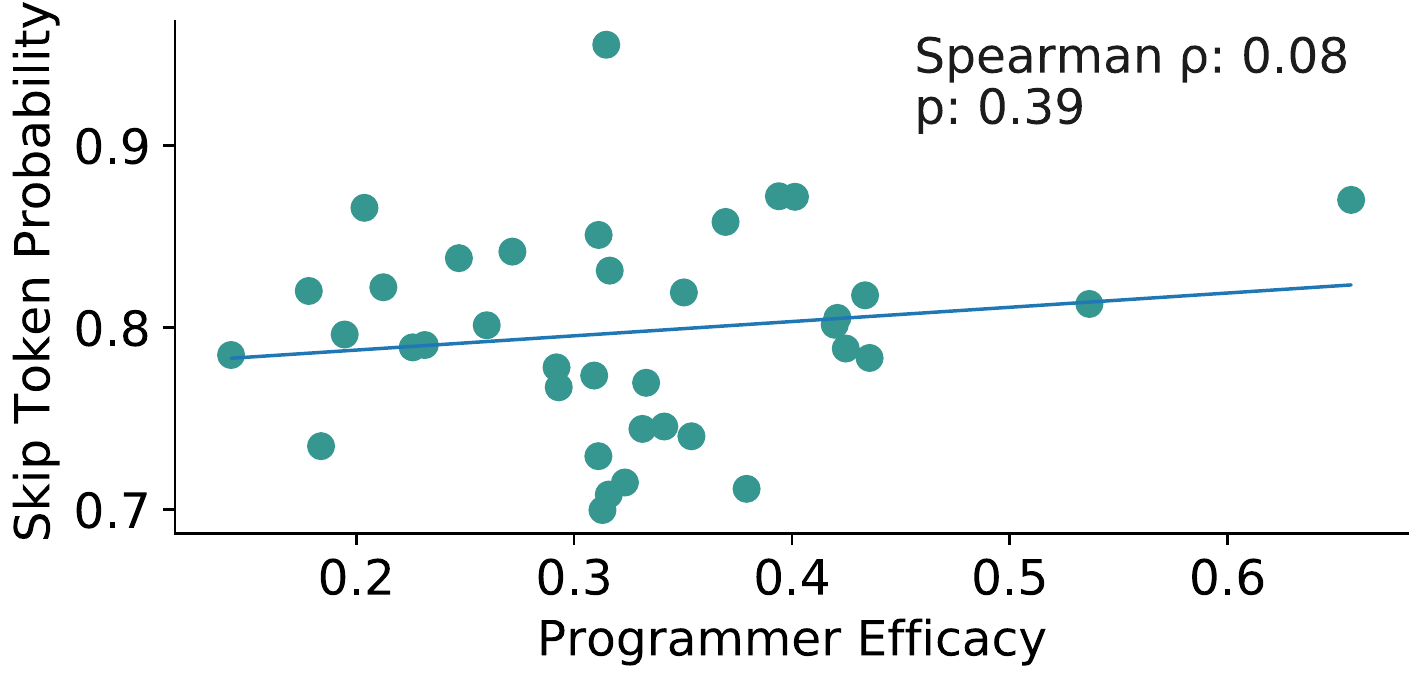}
\hspace{10pt}
\includegraphics[width=0.65\columnwidth]{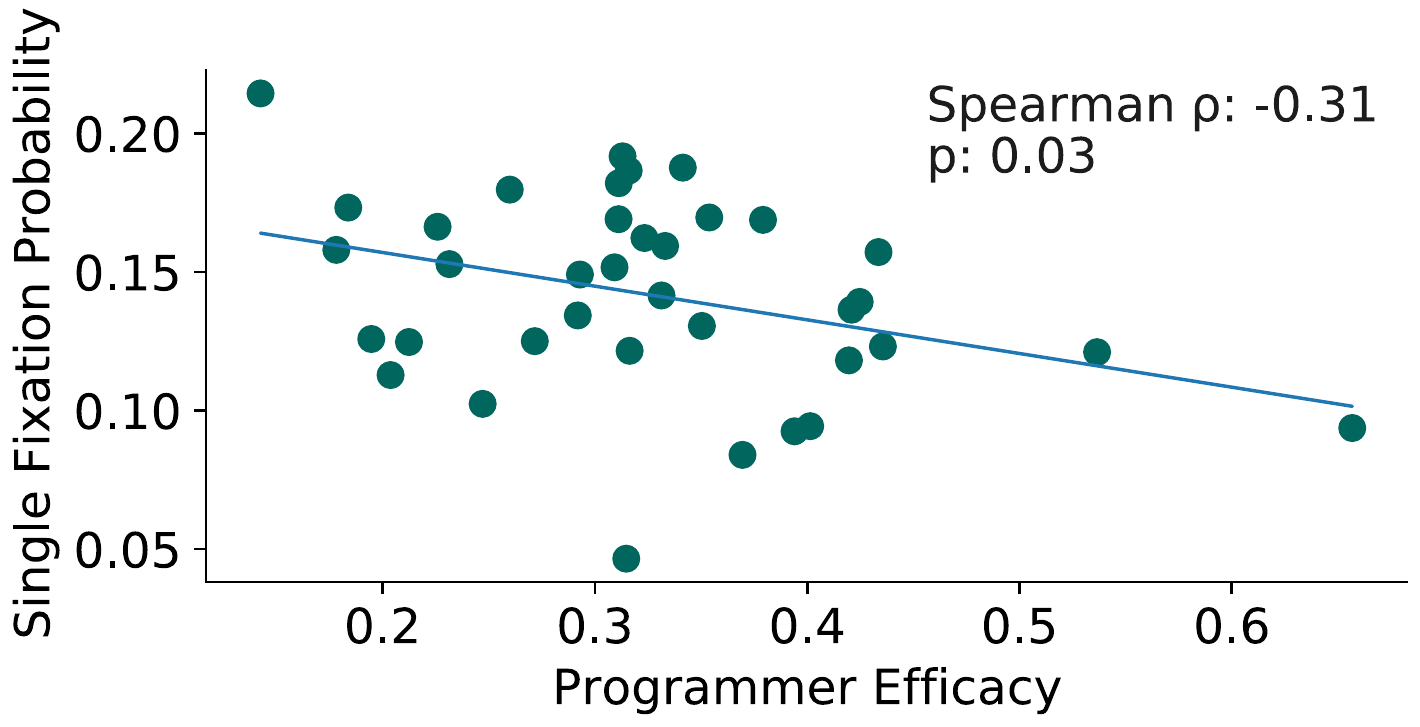}
\hspace{10pt}
\includegraphics[width=0.65\columnwidth]{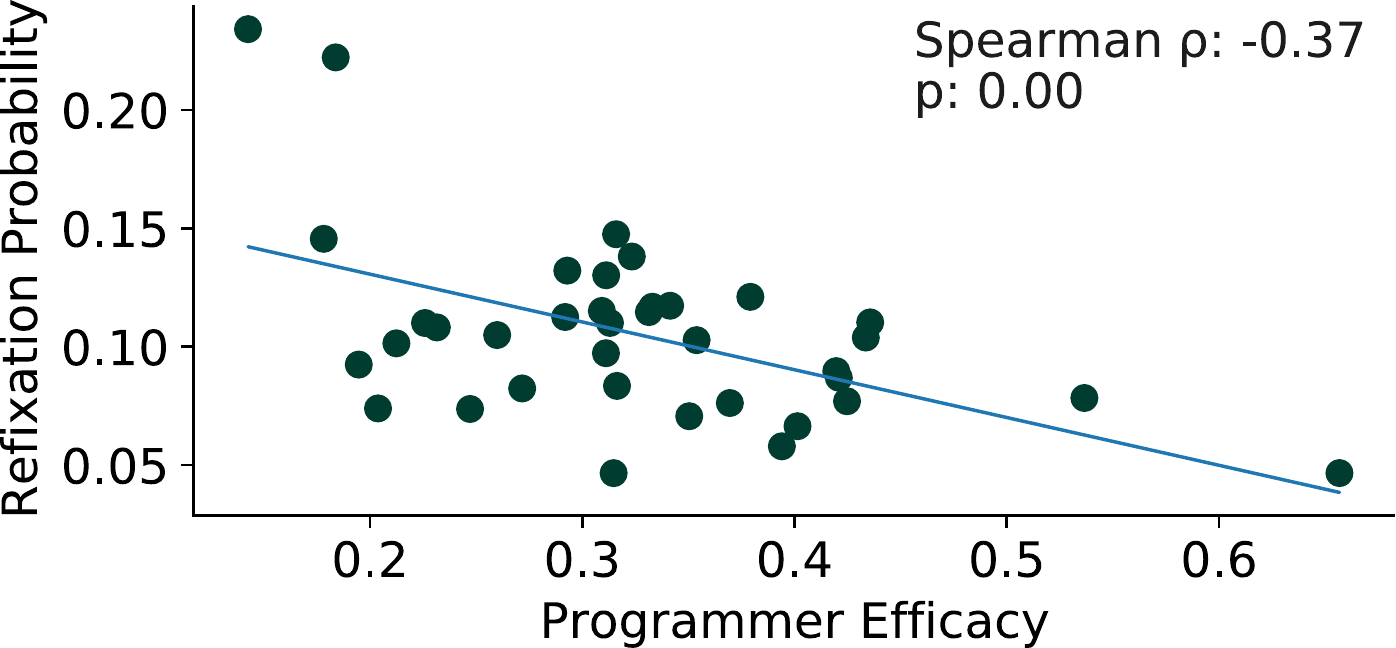}
\caption{Spearman's $\rho$ correlation between programmer efficacy and different eye-tracking measures: Programmers with high efficacy show an efficient reading strategy in terms of shorter fixations, skipping more code elements, and fewer (re)fixations.}
\label{fig:EyetrackingCorrelationPlots}
\end{figure*}

\begin{figure}[h]
    \centering
    \includegraphics[width =\columnwidth]{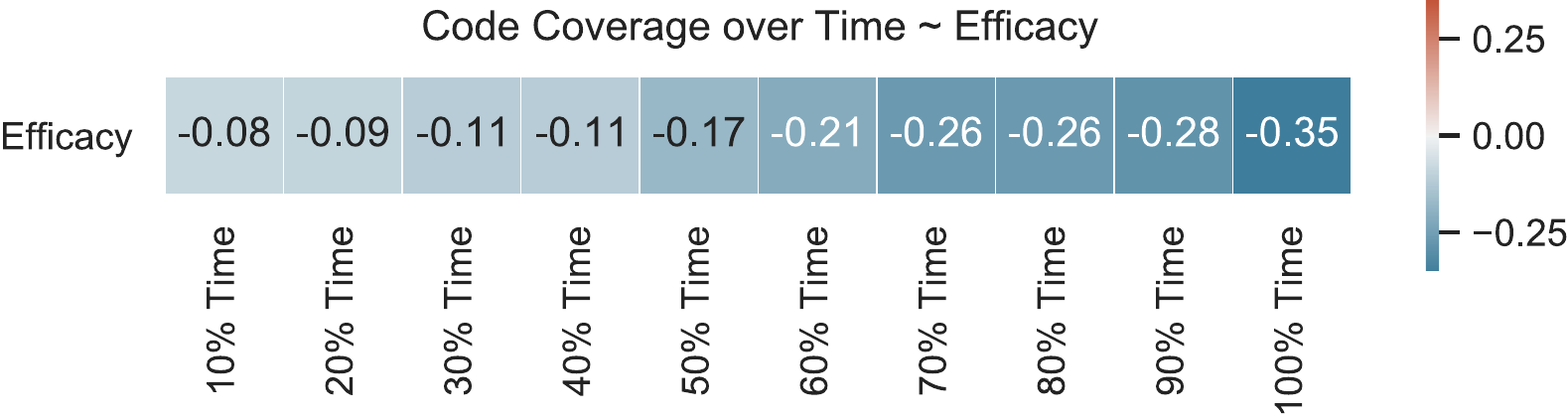}
    \caption{Spearman's $\rho$ correlation between programmer efficacy and code element coverage over time. In the beginning of a task, the difference in code element coverage for different efficacy levels is not substantial. It grows throughout the task in that programmer with high efficacy skip more code elements.}
    \label{fig:CodeCoverageHeatmap}
\end{figure}

\begin{figure*}[ht]
    \centering
    \includegraphics[trim={0 11cm 4.5cm 0}, clip, width =\textwidth]{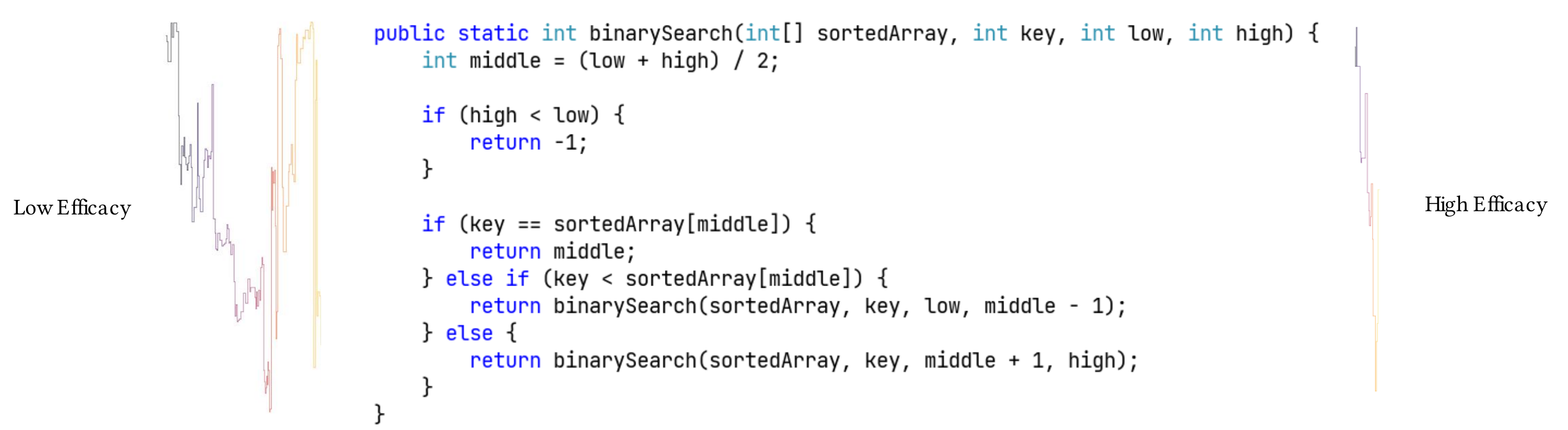}
    \caption{Example scanpath of a programmer with low efficacy (left) and with high efficacy (right). The lines indicate the vertical position in the snippet during comprehension of the code snippet. Clearly, the programmer with high efficacy displays a more targeted, efficient reading strategy.}
    \label{fig:EyetrackingScanpathSample}
\end{figure*}

Our eye-tracking data show a change in reading behavior with increasing efficacy levels. Based on the token-level measures from Al Madi et al., we found that increased efficacy leads to:

\begin{compactitem}
\item shorter (first) fixations ($\rho=-0.14$),
\item shorter gaze duration ($\rho=-0.19$),
\item a much lower chance that a token is revisited ($\rho=-0.37$), and
\item a slightly higher probability that a token is skipped ($\rho=0.08$).
\end{compactitem}
We visualize these relationships in~\Cref{fig:EyetrackingCorrelationPlots}.

Regarding code element coverage, based on Aljehane et al. and Busjahn et al., we find a similar result, such that higher efficacy leads to a lower code element coverage ($\rho=-0.35$). Notably, this difference increases throughout the task, as we illustrate in~\Cref{fig:CodeCoverageHeatmap}. To underline the difference between programmer with low and high efficacy, we show an example scanpath in~\Cref{fig:EyetrackingScanpathSample} (Page~\pageref{fig:EyetrackingScanpathSample}). Clearly, the programmer with high efficacy requires fewer fixations and refixations to comprehend the source-code snippet.

Overall, this leads to the following answer to our research question:

\RQAnswer{RQ\textsubscript{1}}{We can confirm prior results that programmers with higher efficacy read code more efficiently in terms of shorter fixations on fewer code elements. However, some measures show only weak correlations.}

\paragraph{Discussion}

Our results for RQ\textsubscript{1} confirm prior results and corroborate the theory that proficient programmers read source code more efficiently~\cite{Nivala2016} and are actively looking for an efficient way to solve a task~\cite{Roehm2012}. This supports the view that their knowledge guides their eyes~\cite{Vans1999}. However, the established measures that we use capture this effect to different degrees. Fixation duration measures show only weak correlations ($\rho=-0.14$), while fixation probabilities show, at best, medium correlations with programmer efficacy ($\rho=-0.37$). These results support that proficient programmers read individual elements faster, and, in particular, focus on the important elements---skipping several parts that they deem unimportant. This narrow focus on important elements is further highlighted in the code element coverage measure. Programmers with high efficacy read fewer code elements throughout the task, increasing over time as shown in~\Cref{fig:CodeCoverageHeatmap}. Even at the end of a task, they still only fixated on~69\,\% of the tokens, on average---skipping tokens they rate as irrelevant.

While a narrow focus enables high performers to quickly solve a task, they might overlook relevant elements. For example, research in psychology has shown that experts' internal filtering heuristics can lead them to miss relevant information and make mistakes~\cite{Shanteau1992Competence}. In programming, this effect was observed with experts missing obvious syntax errors that novices consistently notice~\cite{Peitek2020:Linearity}.

\subsection{Cognitive Load (RQ\textsubscript{2})}

\begin{table}[h]
\caption{Spearman's $\rho$ correlation between cognitive-load measures and programmer efficacy.}
\begin{tabular}{lr}
\toprule
\makecell{Alpha/Theta\\ Measures of Cognitive Load} & \makecell{Correlation with\\ Programmer Efficacy} \\ \midrule
Ratio\textsubscript{mean} & $-$0.09 \\
Ratio\textsubscript{median} & $-$0.08 \\
Ratio\textsubscript{min} & 0.19 \\
Ratio\textsubscript{max} & $-$0.15 \\
\bottomrule
\end{tabular}
\label{tab:NeuralCorrelates}
\end{table}

\begin{figure*}[ht]
    \centering
    \includegraphics[trim={0 0 2cm 0}, clip, width =1.9\columnwidth]{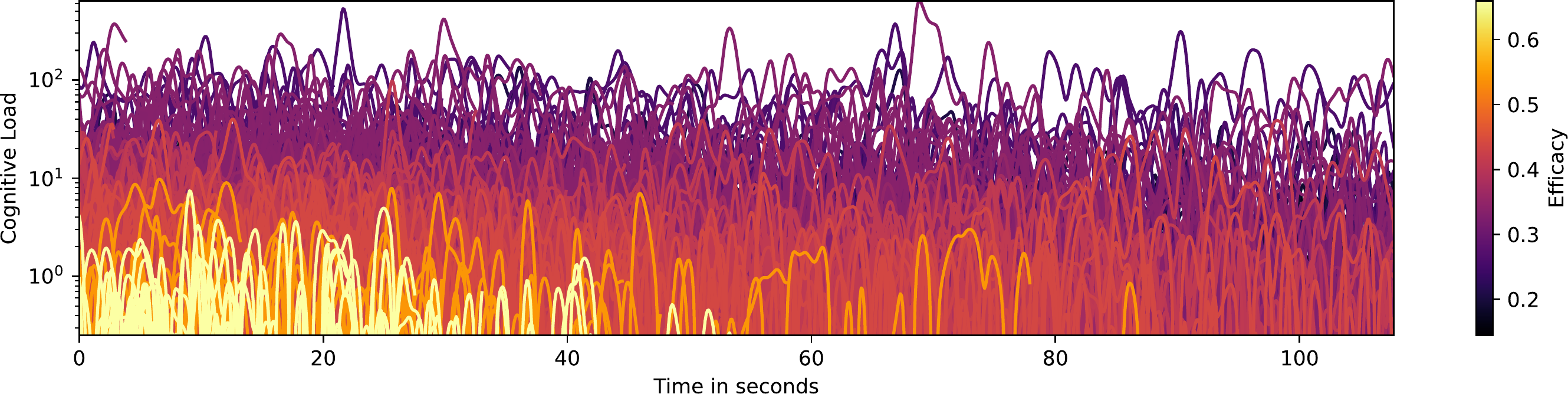}
    \caption{Visualization of cognitive-load measure (i.e., ratio of alpha and theta band) over comprehension-task time. Each line is one participant with the line color indicating their efficacy level.}
    \label{fig:CognitiveLoadOverTime}
\end{figure*}

\noindent
The observed ratio of alpha and theta band as an indicator of cognitive load can be analyzed regarding different features, which we summarize in~\Cref{tab:NeuralCorrelates}. While generally the cognitive load appears to be lower for programmers with higher efficacy ($\rho = -$0.09), the correlations are weak. Still, it has to be taken into account that programmers with higher efficacy tend to complete their tasks faster. In~\Cref{fig:CognitiveLoadOverTime}, we visualize the cognitive load over time depending on efficacy, which shows that the level of cognitive load generally stays lower, and with fewer spikes, for programmers with higher efficacy.

\begin{figure}[h]
    \centering
    \includegraphics[width=\columnwidth]{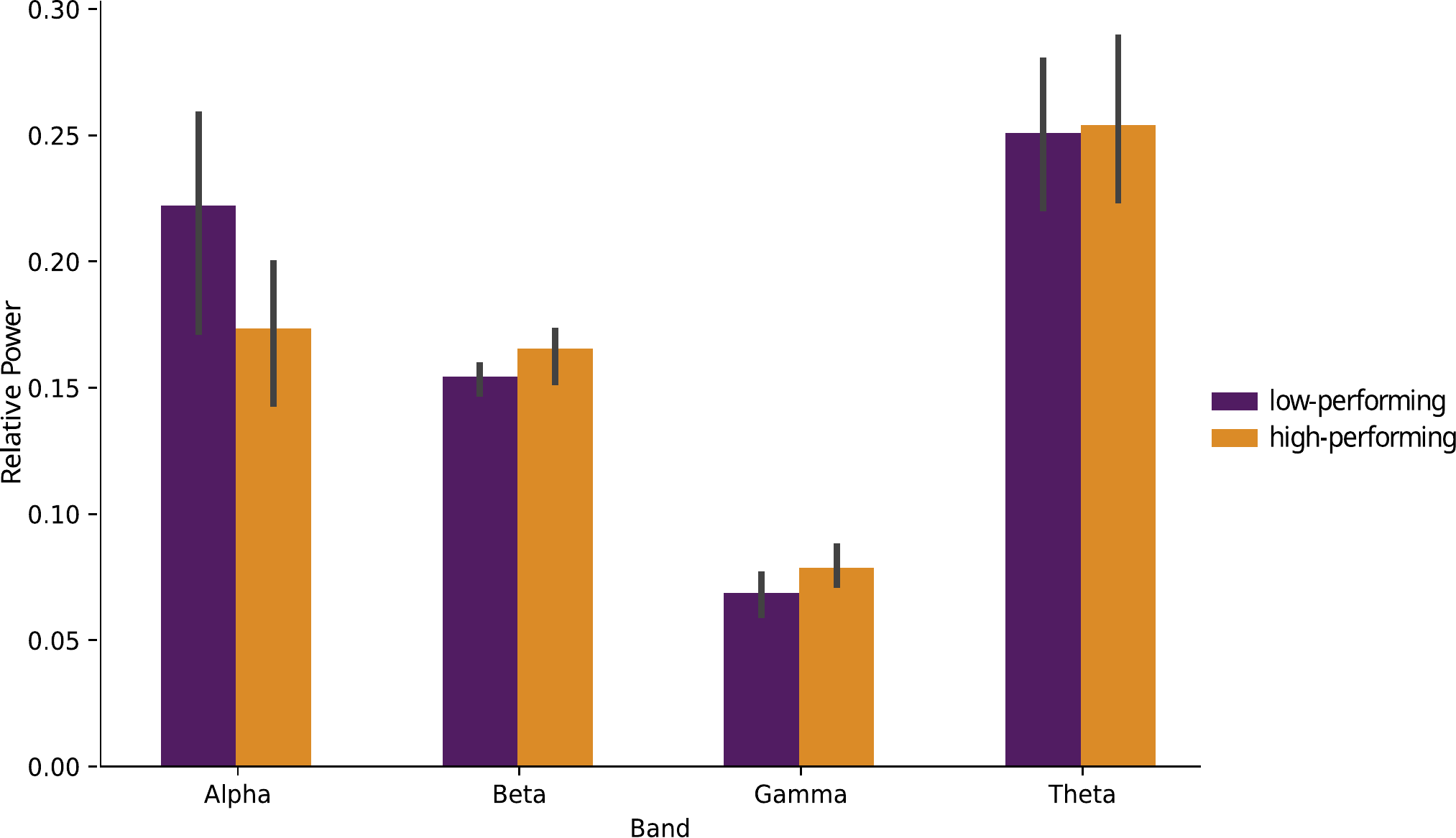}
    \caption{Visualization of relative power along alpha, beta, theta, and gamma bands for the high-performing (efficacy top 33\%) and low-performing group (efficacy bottom 33\%).}
    \label{fig:PowerBands}
\end{figure}

Results from our powerband analysis confirm results from Lee et al.~\cite{Lee2016}. Like Lee et al., we separated our participants into a high-performing and low-performing group~(cf.~\Cref{sec:EEGAnalysis}). We found a lower beta power for programmers with higher efficacy, and higher alpha and gamma power, which is illustrated in~\Cref{fig:PowerBands}.

\RQAnswer{RQ\textsubscript{2}}{The cognitive load is slightly lower for programmers with higher efficacy, despite faster completion times. Programmers with higher efficacy further experience fewer high spikes of cognitive load. This is in line with previous findings of lower beta and higher alpha and gamma power for programmers with higher efficacy.}

\paragraph{Discussion}
The results of RQ\textsubscript{2} indicate that programmers with higher efficacy can not only comprehend source code faster, but also with less mental effort. This could be explained by an increased \emph{neural efficiency}, which has been shown in other fields~\cite{Neubauer2009} or as an effect of source code structures serving as beacons for program comprehension~\cite{Siegmund2017}, but not as a factor of proficiency in programming. One explanation would be a difference in underlying cognitive processing: Experts may see the presented snippets as a task, that is, they have a strategy to solve the problem, and simply need to implement the solution. This is unlike novice programmers, who also need to find a strategy first, then implement the solution, which leads to higher cognitive load and longer response times~\cite{Mead2006}. Due to the lower cognitive-load levels, programmers with high efficacy likely can sustain longer periods of work. By contrast, programmers with lower efficacy are more likely to be overwhelmed by constant spikes of cognitive load (\Cref{fig:CognitiveLoadOverTime}).

\paragraph{Synthesis of RQs\textsubscript{1--2}}
Our study was inspired by several prior experiments investigating program comprehension with various operationalizations and different definitions of experience. We aimed at finding a common ground across their measures, specifically when focusing on programmer efficacy. Overall, our study confirmed the accuracy of some of these measures, but to different degrees. Programmers with higher efficacy can be particularly identified due to their efficient reading strategy and lower spikes in cognitive load. This is not only important for future research, but also has practical implications. For example, a lot of hiring processes use technical interviews in front of a whiteboard, which artificially introduce stress and high cognitive load~\cite{Behroozi2020Debugging}. An alternative solution to evaluate potential talent with measures that allow for accurate and quick responses, without inducing unnecessary stress and cognitive load, would be private interviews, comprehension tasks, or other alternative interview methods, for example, as proposed by Behroozi et al.~\cite{Behroozi2019}.

Regarding programming language design, the combination of eye tracking and EEG may have potential. Future work shall explore whether we can use eye-tracking data to pinpoint spikes in cognitive load to specific code elements, similar to Fakhoury et al.'s work with functional near-infrared spectroscopy (fNIRS)~\cite{Fakhoury2019} or our prior work on combining eye tracking and fMRI~\cite{Peitek2018:Conjoint}. Such combination may reveal code structures that are particularly difficult for programmers to comprehend.

\subsection{Experience Measures (RQ\textsubscript{3})}
\label{sec:ResultsEfficacyExperience}

\begin{table}[h]
\caption{Spearman's $\rho$ correlation between various experience measures and programmer efficacy. Bold text highlights medium and strong correlations ($\rho \geq 0.3$). All measures, including a full correlation matrix, raw data, and results are available on the project's Web site.}

\begin{tabular}{@{}llr@{}}
\toprule
\multicolumn{2}{l}{Measure of Programming Experience} & \makecell{$\rho$ Corr. with\\Efficacy} \\ \midrule
\multirow{4}{*}{\rot{Time}} & Years of Programming & 0.15 \\
 & Years of Professional Programming & 0.14 \\
 & Years of Java & 0.04 \\
 & Years at Work & $-$0.06 \\
\specialrule{0.1pt}{1pt}{1pt}
\multirow{6}{*}{\rot{Self-Estimation}} & Experience Logical Paradigm & 0.07 \\
 & Experience Functional Paradigm & 0.22 \\
 & Experience Object-Oriented Paradigm & 0.17 \\
 & \textbf{Experience Imperative Paradigm} & \textbf{0.32} \\
 & \textbf{Comparison to Peers} & \textbf{0.59} \\
 & \textbf{Comparison to 10-Year Programmer} & \textbf{0.38} \\
\specialrule{0.1pt}{1pt}{1pt}
\multirow{9}{*}{\rot{\makecell{Work Hours per Week\\(Professionals Only, n=26)}}} & \textbf{Overall} & \textbf{0.43} \\
 & Programming & 0.06 \\
 & \textbf{Code Review} & \textbf{0.47} \\
 & Meeting & 0.16 \\
 & \textbf{Tests} & \textbf{0.35} \\
 & Deploy & 0.10 \\
 & \textbf{Mentoring} & \textbf{0.32} \\
 & \textbf{Learning} & \textbf{0.32} \\
 & Other & 0.27 \\
\specialrule{0.1pt}{1pt}{1pt}
 & \textbf{Number of Programming Languages} & \textbf{0.42} \\
\bottomrule
\end{tabular}
\label{tab:AllExpMeasuresCorr}
\end{table}

In~\Cref{tab:AllExpMeasuresCorr}, we show the strength of correlations between programmer efficacy and a subset of popular measures of programming experience. Notably, some commonly used experience measures (e.g., years of (professional) programming~\cite{Bednarik2012}) show little predictive power to our participants' efficacy. But, several other dimensions of experience show, at least, medium-strength correlations. Specifically, self-estimation in comparison to peers ($\rho=0.59$) and a 10-year professional programmer ($\rho=0.38$) show that our participants seem to be keenly aware of their proficiency level. The number of known programming languages also shows a medium-strength correlation ($\rho=0.42$). For the subset of professional programmers~($n=26$), several questions that capture learning eagerness (e.g., learning, mentoring, code review) show medium correlations. However, strikingly, the time professionals spend on pure programming does not correlate to their performance in our experiment ($\rho=0.06$).

\newpage
Overall, these results allow us to answer this research question:

\RQAnswer{RQ\textsubscript{3}}{Programmer efficacy does not correlate with commonly used experience measures, such as years of programming. Self-estimation and indicators of learning eagerness show, at least, medium correlations with observed programmer efficacy.}

\paragraph{Discussion}
\label{sec:EfficacyExperienceDiscussion}

Our experiment highlights two fundamental problems for studying programmers. One issue is that programming is such a diverse field with different technologies that require different skill sets, so it is incredibly difficult to accurately measure a programmer's experience. Therefore, many researchers rely on simple measures, such as years of programming. However, this can become problematic if the difference between the selected measure and the \emph{actual} proficiency level becomes a significant confounding factor~\cite{Siegmund2015Confounding}. Our experiment, in line with prior research~\cite{Dieste2017}, underlines how limited common experience measures are in predicting programmers' efficacy. Thus, future research must carefully consider collecting more comprehensive experience data, in particular, when using it to separate the participants into groups. While this may not be completely new insights, our experiment further corroborates this critical point. The use of too simplistic measures can potentially weaken empirical studies of programmers and their conclusions.

\section{Threats to Validity}
\label{sec:threats}

In this section, we discuss threats to construct, internal, and external validity of our study.

\subsection{Construct Validity}

Our programmer efficacy measure may deviate from a participant's \emph{true} efficacy due to the nature of our selected snippets and programming language. We mitigated this threat by selecting a variety of source-code snippets such that prior knowledge plays a minuscule role, and we ensured sufficient Java knowledge for each participant.

\subsection{Internal Validity}
\label{sec:threatsInternal}

We have operationalized program comprehension in a multiple-step design, in which participants first had to comprehend a source-code snippet, then compute an input/output task, and finally select the output from four answers. Clearly, program comprehension is a multi-faceted phenomenon for which a variety of operationalizations are possible~\cite{Dunsmore2000}, but our approach ensures that participants genuinely comprehended each snippet. 

Regarding data collection, we limited the experiment to a maximum of 1~hour (with two breaks) to avoid fatigue effects. Depending on each participant's speed, this led to an unequal number of collected data points (17 of 37~participants completed all 32~snippets). While we could have ended the experiment after, for example, 25 instead of 32~snippets for everyone, we would have lost around 10\% of data. Therefore, we chose a small potential bias as a trade-off for a notably larger data set (to gain more statistical power and external validity). We mitigated the threat of an unequal number of snippets by randomizing the presentation order of snippets.

Our results indicate a strong influence regarding measures of several programming activities, including the amount of code review and testing, on observed efficacy. This relationship between observed efficacy and time spent with these identified programming activities may be overly emphasized by our experiment design since they are close in nature to our program-comprehension task.

\subsection{External Validity}
Due to our focus on internal validity, our presented study is limited regarding its generalizability. While our participant sample covers a wide range of experience and proficiency levels, the presented tasks were limited to algorithms in an object-oriented programming language. Thus, our study on program comprehension may not generalize to all software-engineering activities, including comprehension of large code bases and other programming paradigms. Still, our study pinned down an effect with high internal validity; future work shall replicate, vary, and extend the setup toward external validity.

\section{Related Work}

Our study is at the intersection between the fields of program comprehension, eye tracking, neuroimaging, and programming expertise, which we each discuss below.

\paragraph{Program Comprehension}
There is a multitude of program-com-\\prehension studies~\cite{Soloway1984, Ajami2017, Busjahn2015, Siegmund2014:fMRI, Lee2016, Fakhoury2018}, on which we build by re-using source-code snippets as a foundation for our experiment. In the early years, behavioral or reflective studies were popular~\cite{Siegmund2016}. Our study shares a similar design with them in terms of snippets and tasks, but our observed measures are more in line with the recent movement toward more objective measurement methods, such as eye tracking or EEG~\cite{Siegmund2016}, which we discuss next.

\paragraph{Eye Tracking}
Eye tracking enables researchers to objectively observe attention during complex tasks and has increased in popularity, including program comprehension~\cite{Sharafi2015Systematic}. Again, there is a multitude of studies that use various eye-gaze measures~\cite{Crosby1990, Bednarik2012, Busjahn2015, Nivala2016, Aljehane2021, AlMadi2021, Peachock2017, Sharif2012, Blascheck2019, Bauer2019, Peitek2020:Linearity}. We have used recently established eye-gaze measures for differentiation of programmer expertise~\cite{Aljehane2021, AlMadi2021, Busjahn2015}, but we applied these measures to a larger pool of diverse programmers than before and included insights from neural correlates via EEG data. With efficacy, we also arguably used a better way to separate participants~(cf.~\Cref{sec:EfficacyExperienceDiscussion}).

\paragraph{Neural Correlates of Program Comprehension}
In addition to eye tracking, some studies have employed methods to observe neural correlates of program comprehension. Some have used fMRI~\cite{Floyd2017, Castelhano2018} or fNIRS~\cite{Nakagawa2014, Fakhoury2019}, others have used EEG~\cite{Lee2016, Yeh2017, Kosti2018, Medeiros2019EEG, Medeiros2021, Lin2021}. Our study adopted the measures of cognitive load of Lee et al. (i.e., alpha and beta power)~\cite{Lee2016} and Medeiros et al. (i.e., ratio of theta and alpha)~\cite{Medeiros2021}, but it differs in terms of the goal and the participant pool. Lee et al. focused on classification between two distinct groups, while Medeiros et al. focused on software metrics~\cite{Medeiros2021}. By contrast, we focused on the programmer and investigated their efficacy on a continuous distribution, without the need to create two groups.

\paragraph{Programming Experience and Expertise}
There have been multiple waves of program-comprehension research regarding expertise. In the early years, several studies have devised theories of expertise, in particular related to plans~\cite{Gilmore1988}, mental models~\cite{Wiedenbeck1993, Pennington1987}, strategy~\cite{Koenemann1991} and knowledge~\cite{Gilmore1990}. More recently, researchers have focused on visual attention~\cite{Bednarik2012} and implementation style~\cite{Ortin2020}. Our study differs in that we operationalize on a more narrow scope of expertise based on programmer efficacy, rather than measures of knowledge, mental representations, or other intangible factors. We argue our operationalization offers a higher relevance to practical and educational problems as well as a clearer definition over an obfuscated experience measure, because of its better link to task performance.

\vspace{\baselineskip}
In summary, our experiment is a fusion of established research designs from different fields and is novel by integrating several measures in one experiment. Furthermore, we focused on inviting a diverse participant pool as well as on efficacy as a separating factor (while still collecting a wide array of experience measures).

\section{Perspectives and Conclusion}
\label{sec:ImplicationsConclusion}

\paragraph{Measuring Programming Experience}
\label{sec:ImplicationsExperience}

For future studies, one major obstacle is to select the correct measure of programming experience. Cognitive psychology has investigated the relationship between experience and expertise in many fields beyond programming~\cite{Shanteau1992}. One consistent finding is that the length of experience can be part of expertise, as people continue to acquire skill~\cite{Dreyfus1986}, but is not everything~\cite{Shanteau1992Competence}. In programming, studies from the 1990s identified a similar effect, in that there are two elements at play: the time and learning~\cite{Stanislaw1994}. In our questionnaire, we aimed to capture many aspects of programming experience, including measures of learning, but also regarding content creation and work-time distribution. One theme that consistently showed a strong connection with observed efficacy is the eagerness of learning and inquisitiveness. The number of known programming languages, how much time is spent on mentoring, and on code review are all highly relevant factors. Psychology describes this conscious effort to keep learning as \emph{deliberate practice}. It also highlights that not necessarily the length of practice, but the intensity and goal-orientation is critical to achieve expertise in a topic~\cite{Ericsson1996}. The effect of deliberate practice has been observed in programming education and practice, as well. For example, students show higher ability when participating in programming competitions~\cite{Zha2021}. In an industry survey, software engineers shared the notion that a great software engineer is shown by their open-mindedness and eagerness to learn. They considered the ability to learn as more important than technical skill~\cite{Li2015}. This was later confirmed in an experimental setting, in which company seniority level showed little correlation to actual programming skill~\cite{Jorgensen2020}. 
Overall, the results of our experiment underlined that the learning component is a highly relevant measure to capture efficacy, which leads us to suggest it as important measure(s) for future experiments.

\paragraph{Implications for Research}
Beyond learning measures, our results indicate that future research could maximize its potential when collecting self-estimation measures. This result is in line with a similar finding by Siegmund et al., which also highlighted self-estimated comparison as a differentiating factor among students~\cite{Siegmund2014:Exp}. In the same vein, Kleinschmager and Hanenberg found self-estimated comparison, at least, as effective as pretests or university grades in its predictive power of efficacy~\cite{Kleinschmager2011}. Our shared experience questionnaire provides a proven basis for others to use to capture several dimensions of programming experience.

\paragraph{Implications for Industry}
From a practical perspective, our empirical results provide scientific evidence that technical interviews may be necessary for establishing efficacy, since common measures selected for decisions (e.g., years of experience), have limited to no predictive power. Our questionnaire provides a few promising directions for operationalizing a more effective method for assessing efficacy in industry, such as peer review and mentoring experience, peer comparison, and learning behaviors. However, in high-stake decisions, such as promotion or hiring, the degree to which self-estimation can be reliable remains an open question.

\paragraph{Conclusion}
Experience, expertise, and efficacy are three dimensions of characteristics of programmers that are not well understood. In particular, the relationship between them is unclear. In this paper, we have presented correlates of efficacy in terms of reading order and cognitive load across a wide range of programmers. 

Commonly used experience measures do not correlate well to observed efficacy. Instead, we underline to use self-estimation and learning eagerness as more accurate measures for programming experience.

Despite encouraging results, future work shall explore programmer efficacy in more detail. For example, the link between reading behavior and cognitive load could be explored to understand causalities. Is the more efficient reading strategy of programmers with high efficacy the cause for lower cognitive load, or vice versa?

\begin{acks}
We thank all participants of our study. Furthermore, we thank Julia Hess, Tobias Jungbluth, and Johannes Ihl for their support during data acquisition.

Rekrut's work is supported by the German Federal Ministry of Education and Research (01IS12050). Parnin's work is supported by the National Science Foundation under grant number~2045272. Siegmund's work is supported by DFG grant SI~2045/2\=/2. Apel's work is supported by ERC Advanced Grant 101052182 as well as DFG Grant 389792660 as part of TRR 248 -- CPEC.
\end{acks}

\bibliographystyle{ACM-Reference-Format} 
\bibliography{References}


\begin{thebibliography}{81}


\ifx \showCODEN    \undefined \def \showCODEN     #1{\unskip}     \fi
\ifx \showDOI      \undefined \def \showDOI       #1{#1}\fi
\ifx \showISBNx    \undefined \def \showISBNx     #1{\unskip}     \fi
\ifx \showISBNxiii \undefined \def \showISBNxiii  #1{\unskip}     \fi
\ifx \showISSN     \undefined \def \showISSN      #1{\unskip}     \fi
\ifx \showLCCN     \undefined \def \showLCCN      #1{\unskip}     \fi
\ifx \shownote     \undefined \def \shownote      #1{#1}          \fi
\ifx \showarticletitle \undefined \def \showarticletitle #1{#1}   \fi
\ifx \showURL      \undefined \def \showURL       {\relax}        \fi
\providecommand\bibfield[2]{#2}
\providecommand\bibinfo[2]{#2}
\providecommand\natexlab[1]{#1}
\providecommand\showeprint[2][]{arXiv:#2}

\bibitem[\protect\citeauthoryear{??}{myt}{2020}]%
        {myth2020}
 \bibinfo{year}{2020}\natexlab{}.
\newblock \bibinfo{title}{{The Myth of the Developer That Can't Code}}.
\newblock
\newblock
\urldef\tempurl%
\url{https://news.ycombinator.com/item?id=22862871}
\showURL{%
\tempurl}


\bibitem[\protect\citeauthoryear{??}{hac}{2021}]%
        {hackerNews}
 \bibinfo{year}{2021}\natexlab{}.
\newblock \bibinfo{title}{{Facebook Senior Software Engineer Interview}}.
\newblock
\newblock
\urldef\tempurl%
\url{https://news.ycombinator.com/item?id=25658098}
\showURL{%
\tempurl}


\bibitem[\protect\citeauthoryear{Ajami, Woodbridge, and Feitelson}{Ajami
  et~al\mbox{.}}{2017}]%
        {Ajami2017}
\bibfield{author}{\bibinfo{person}{Shulamyt Ajami}, \bibinfo{person}{Yonatan
  Woodbridge}, {and} \bibinfo{person}{Dror Feitelson}.}
  \bibinfo{year}{2017}\natexlab{}.
\newblock \showarticletitle{{Syntax, Predicates, Idioms - What Really Affects
  Code Complexity?}}. In \bibinfo{booktitle}{\emph{Proc.\ Int'l Conf.\ Program
  Comprehension (ICPC)}}. \bibinfo{publisher}{Springer},
  \bibinfo{pages}{66--76}.
\newblock


\bibitem[\protect\citeauthoryear{Al~Madi, Peterson, Sharif, and
  Maletic}{Al~Madi et~al\mbox{.}}{2021}]%
        {AlMadi2021}
\bibfield{author}{\bibinfo{person}{Naser Al~Madi}, \bibinfo{person}{Cole
  Peterson}, \bibinfo{person}{Bonita Sharif}, {and} \bibinfo{person}{Jonathan
  Maletic}.} \bibinfo{year}{2021}\natexlab{}.
\newblock \showarticletitle{{From Novice to Expert: Analysis of Token Level
  Effects in a Longitudinal Eye Tracking Study}}. In
  \bibinfo{booktitle}{\emph{Proc.\ Int'l Conf.\ Program Comprehension (ICPC)}}.
  \bibinfo{publisher}{IEEE}, \bibinfo{pages}{172--183}.
\newblock


\bibitem[\protect\citeauthoryear{Aljehane, Sharif, and Maletic}{Aljehane
  et~al\mbox{.}}{2021}]%
        {Aljehane2021}
\bibfield{author}{\bibinfo{person}{Salwa Aljehane}, \bibinfo{person}{Bonita
  Sharif}, {and} \bibinfo{person}{Jonathan Maletic}.}
  \bibinfo{year}{2021}\natexlab{}.
\newblock \showarticletitle{{Determining Differences in Reading Behavior
  Between Experts and Novices by Investigating Eye Movement on Source Code
  Constructs During a Bug Fixing Task}}. In \bibinfo{booktitle}{\emph{{Proc.\
  Symposium on Eye Tracking Research \& Applications (ETRA)}}}.
  \bibinfo{publisher}{ACM}, Article \bibinfo{articleno}{30},
  \bibinfo{numpages}{6}~pages.
\newblock


\bibitem[\protect\citeauthoryear{Atwood}{Atwood}{2007}]%
        {atwood2022}
\bibfield{author}{\bibinfo{person}{Jeff Atwood}.}
  \bibinfo{year}{2007}\natexlab{}.
\newblock \bibinfo{title}{{Why Can't Programmers... Program?}}
\newblock
\newblock
\urldef\tempurl%
\url{https://blog.codinghorror.com/why-cant-programmers-program/}
\showURL{%
\tempurl}


\bibitem[\protect\citeauthoryear{Bauer, Siegmund, Peitek, Hofmeister, and
  Apel}{Bauer et~al\mbox{.}}{2019}]%
        {Bauer2019}
\bibfield{author}{\bibinfo{person}{Jennifer Bauer}, \bibinfo{person}{Janet
  Siegmund}, \bibinfo{person}{Norman Peitek}, \bibinfo{person}{Johannes
  Hofmeister}, {and} \bibinfo{person}{Sven Apel}.}
  \bibinfo{year}{2019}\natexlab{}.
\newblock \showarticletitle{{Indentation: Simply a Matter of Style or Support
  for Program Comprehension?}}
\newblock In \bibinfo{booktitle}{\emph{Proc.\ Int'l Conf.\ Program
  Comprehension (ICPC)}}. \bibinfo{publisher}{ACM}, \bibinfo{pages}{11}.
\newblock


\bibitem[\protect\citeauthoryear{Bednarik}{Bednarik}{2012}]%
        {Bednarik2012}
\bibfield{author}{\bibinfo{person}{Roman Bednarik}.}
  \bibinfo{year}{2012}\natexlab{}.
\newblock \showarticletitle{{Expertise-Dependent Visual Attention Strategies
  Develop over Time during Debugging with Multiple Code Representations}}.
\newblock \bibinfo{journal}{\emph{Int'l Journal of Human-Computer Studies}}
  \bibinfo{volume}{70}, \bibinfo{number}{2} (\bibinfo{year}{2012}),
  \bibinfo{pages}{143--155}.
\newblock


\bibitem[\protect\citeauthoryear{Behroozi, Parnin, and Barik}{Behroozi
  et~al\mbox{.}}{2019}]%
        {Behroozi2019}
\bibfield{author}{\bibinfo{person}{Mahnaz Behroozi}, \bibinfo{person}{Chris
  Parnin}, {and} \bibinfo{person}{Titus Barik}.}
  \bibinfo{year}{2019}\natexlab{}.
\newblock \showarticletitle{{Hiring Is Broken: What Do Developers Say About
  Technical Interviews?}}. In \bibinfo{booktitle}{\emph{Symposium on Visual
  Languages and Human-Centric Computing (VL/HCC)}}. \bibinfo{publisher}{IEEE},
  \bibinfo{pages}{1--9}.
\newblock


\bibitem[\protect\citeauthoryear{Behroozi, Shirolkar, Barik, and
  Parnin}{Behroozi et~al\mbox{.}}{2020}]%
        {Behroozi2020Debugging}
\bibfield{author}{\bibinfo{person}{Mahnaz Behroozi}, \bibinfo{person}{Shivani
  Shirolkar}, \bibinfo{person}{Titus Barik}, {and} \bibinfo{person}{Chris
  Parnin}.} \bibinfo{year}{2020}\natexlab{}.
\newblock \showarticletitle{{Debugging Hiring: What Went Right and What Went
  Wrong in the Technical Interview Process}}. In
  \bibinfo{booktitle}{\emph{Proc.\ Int'l Conf.\ Software Engineering: Software
  Engineering in Society (ICSE-SEIS)}}. \bibinfo{publisher}{IEEE},
  \bibinfo{pages}{71--80}.
\newblock


\bibitem[\protect\citeauthoryear{Blascheck and Sharif}{Blascheck and
  Sharif}{2019}]%
        {Blascheck2019}
\bibfield{author}{\bibinfo{person}{Tanja Blascheck} {and}
  \bibinfo{person}{Bonita Sharif}.} \bibinfo{year}{2019}\natexlab{}.
\newblock \showarticletitle{{Visually Analyzing Eye Movements on Natural
  Language Texts and Source Code Snippets}}. In
  \bibinfo{booktitle}{\emph{{Proc.\ Symposium on Eye Tracking Research \&
  Applications (ETRA)}}}. \bibinfo{publisher}{ACM},
  \bibinfo{pages}{14:1--14:9}.
\newblock


\bibitem[\protect\citeauthoryear{Burkhardt, D{\'e}tienne, and
  Wiedenbeck}{Burkhardt et~al\mbox{.}}{2002}]%
        {Burkhardt2002}
\bibfield{author}{\bibinfo{person}{Jean-Marie Burkhardt},
  \bibinfo{person}{Fran{\c{c}}oise D{\'e}tienne}, {and} \bibinfo{person}{Susan
  Wiedenbeck}.} \bibinfo{year}{2002}\natexlab{}.
\newblock \showarticletitle{{Object-Oriented Program Comprehension: Effect of
  Expertise, Task and Phase}}.
\newblock \bibinfo{journal}{\emph{Empirical Softw.\ Eng.}} \bibinfo{volume}{7},
  \bibinfo{number}{2} (\bibinfo{year}{2002}), \bibinfo{pages}{115--156}.
\newblock


\bibitem[\protect\citeauthoryear{Busjahn, Bednarik, Begel, Crosby, Paterson,
  Schulte, Sharif, and Tamm}{Busjahn et~al\mbox{.}}{2015}]%
        {Busjahn2015}
\bibfield{author}{\bibinfo{person}{Teresa Busjahn}, \bibinfo{person}{Roman
  Bednarik}, \bibinfo{person}{Andrew Begel}, \bibinfo{person}{Martha Crosby},
  \bibinfo{person}{James Paterson}, \bibinfo{person}{Carsten Schulte},
  \bibinfo{person}{Bonita Sharif}, {and} \bibinfo{person}{Sascha Tamm}.}
  \bibinfo{year}{2015}\natexlab{}.
\newblock \showarticletitle{{Eye Movements in Code Reading: Relaxing the Linear
  Order}}. In \bibinfo{booktitle}{\emph{Proc.\ Int'l Conf.\ Program
  Comprehension (ICPC)}}. \bibinfo{publisher}{IEEE}, \bibinfo{pages}{255--265}.
\newblock


\bibitem[\protect\citeauthoryear{Castelhano, Duarte, Ferreira, Dur{\~a}es,
  Madeira, and Castelo-Branco}{Castelhano et~al\mbox{.}}{2019}]%
        {Castelhano2018}
\bibfield{author}{\bibinfo{person}{Jo{\~a}o Castelhano},
  \bibinfo{person}{Isabel Duarte}, \bibinfo{person}{Carlos Ferreira},
  \bibinfo{person}{Jo{\~a}o Dur{\~a}es}, \bibinfo{person}{Henrique Madeira},
  {and} \bibinfo{person}{Miguel Castelo-Branco}.}
  \bibinfo{year}{2019}\natexlab{}.
\newblock \showarticletitle{{The Role of the Insula in Intuitive Expert Bug
  Detection in Computer Code: An fMRI Study}}.
\newblock \bibinfo{journal}{\emph{Brain Imaging and Behavior}}
  \bibinfo{volume}{13}, \bibinfo{number}{3} (\bibinfo{year}{2019}),
  \bibinfo{pages}{623--637}.
\newblock


\bibitem[\protect\citeauthoryear{Charness, Gneezy, and Kuhn}{Charness
  et~al\mbox{.}}{2012}]%
        {Charness2012Experimental}
\bibfield{author}{\bibinfo{person}{Gary Charness}, \bibinfo{person}{Uri
  Gneezy}, {and} \bibinfo{person}{Michael Kuhn}.}
  \bibinfo{year}{2012}\natexlab{}.
\newblock \showarticletitle{Experimental Methods: Between-Subject and
  Within-Subject Design}.
\newblock \bibinfo{journal}{\emph{Journal of Economic Behavior \&
  Organization}} \bibinfo{volume}{81}, \bibinfo{number}{1}
  (\bibinfo{year}{2012}), \bibinfo{pages}{1--8}.
\newblock


\bibitem[\protect\citeauthoryear{Crk and Kluthe}{Crk and Kluthe}{2014}]%
        {Crk2014}
\bibfield{author}{\bibinfo{person}{Igor Crk} {and} \bibinfo{person}{Timothy
  Kluthe}.} \bibinfo{year}{2014}\natexlab{}.
\newblock \showarticletitle{{Toward Using Alpha and Theta Brain Waves to
  Quantify Programmer Expertise}}. In \bibinfo{booktitle}{\emph{Int'l Conf.
  Engineering in Medicine and Biology Society}}. \bibinfo{publisher}{IEEE},
  \bibinfo{pages}{5373--5376}.
\newblock


\bibitem[\protect\citeauthoryear{Crosby, Scholtz, and Wiedenbeck}{Crosby
  et~al\mbox{.}}{2002}]%
        {Crosby2002}
\bibfield{author}{\bibinfo{person}{Martha Crosby}, \bibinfo{person}{Jean
  Scholtz}, {and} \bibinfo{person}{Susan Wiedenbeck}.}
  \bibinfo{year}{2002}\natexlab{}.
\newblock \showarticletitle{{The Roles Beacons Play in Comprehension for Novice
  and Expert Programmers}}. In \bibinfo{booktitle}{\emph{Annual Conf.\
  Psychology of Programming Interest Group (PPIG)}}. \bibinfo{pages}{58--73}.
\newblock


\bibitem[\protect\citeauthoryear{Crosby and Stelovsky}{Crosby and
  Stelovsky}{1990}]%
        {Crosby1990}
\bibfield{author}{\bibinfo{person}{Martha Crosby} {and} \bibinfo{person}{Jan
  Stelovsky}.} \bibinfo{year}{1990}\natexlab{}.
\newblock \showarticletitle{{How Do We Read Algorithms? A Case Study}}.
\newblock \bibinfo{journal}{\emph{Computer}} \bibinfo{volume}{23},
  \bibinfo{number}{1} (\bibinfo{year}{1990}), \bibinfo{pages}{25--35}.
\newblock


\bibitem[\protect\citeauthoryear{Delorme and Makeig}{Delorme and
  Makeig}{2004}]%
        {Delorme2004}
\bibfield{author}{\bibinfo{person}{Arnaud Delorme} {and} \bibinfo{person}{Scott
  Makeig}.} \bibinfo{year}{2004}\natexlab{}.
\newblock \showarticletitle{{EEGLAB: An Open Source Toolbox for Analysis of
  Single-Trial EEG Dynamics Including Independent Component Analysis}}.
\newblock \bibinfo{journal}{\emph{Journal of Neuroscience Methods}}
  \bibinfo{volume}{134}, \bibinfo{number}{1} (\bibinfo{year}{2004}),
  \bibinfo{pages}{9--21}.
\newblock


\bibitem[\protect\citeauthoryear{Dieste, Aranda, Uyaguari, Turhan, Tosun,
  Fucci, Oivo, and Juristo}{Dieste et~al\mbox{.}}{2017}]%
        {Dieste2017}
\bibfield{author}{\bibinfo{person}{Oscar Dieste}, \bibinfo{person}{Alejandrina
  Aranda}, \bibinfo{person}{Fernando Uyaguari}, \bibinfo{person}{Burak Turhan},
  \bibinfo{person}{Ayse Tosun}, \bibinfo{person}{Davide Fucci},
  \bibinfo{person}{Markku Oivo}, {and} \bibinfo{person}{Natalia Juristo}.}
  \bibinfo{year}{2017}\natexlab{}.
\newblock \showarticletitle{{Empirical Evaluation of the Effects of Experience
  on Code Quality and Programmer Productivity: An Exploratory Study}}.
\newblock \bibinfo{journal}{\emph{Empirical Softw.\ Eng.}}
  \bibinfo{volume}{22}, \bibinfo{number}{5} (\bibinfo{year}{2017}),
  \bibinfo{pages}{2457--2542}.
\newblock


\bibitem[\protect\citeauthoryear{Dreyfus and Dreyfus}{Dreyfus and
  Dreyfus}{1986}]%
        {Dreyfus1986}
\bibfield{author}{\bibinfo{person}{Hubert Dreyfus} {and}
  \bibinfo{person}{Stuart Dreyfus}.} \bibinfo{year}{1986}\natexlab{}.
\newblock \bibinfo{booktitle}{\emph{{Mind over Machine: The Power of Human
  Intuition and Expertise in the Era of the Computer}}}.
\newblock


\bibitem[\protect\citeauthoryear{Dunsmore and Roper}{Dunsmore and
  Roper}{2000}]%
        {Dunsmore2000}
\bibfield{author}{\bibinfo{person}{Alastair Dunsmore} {and}
  \bibinfo{person}{Marc Roper}.} \bibinfo{year}{2000}\natexlab{}.
\newblock \bibinfo{booktitle}{\emph{{A Comparative Evaluation of Program
  Comprehension Measures}}}.
\newblock \bibinfo{type}{{T}echnical {R}eport} EFoCS 35-2000.
  \bibinfo{institution}{Department of Computer Science, University of
  Strathclyde}.
\newblock


\bibitem[\protect\citeauthoryear{Ericsson and Lehmann}{Ericsson and
  Lehmann}{1996}]%
        {Ericsson1996}
\bibfield{author}{\bibinfo{person}{Anders Ericsson} {and}
  \bibinfo{person}{Andreas Lehmann}.} \bibinfo{year}{1996}\natexlab{}.
\newblock \showarticletitle{{Expert and Exceptional Performance: Evidence of
  Maximal Adaptation to Task Constraints}}.
\newblock \bibinfo{journal}{\emph{Annual Review of Psychology}}
  \bibinfo{volume}{47}, \bibinfo{number}{1} (\bibinfo{year}{1996}),
  \bibinfo{pages}{273--305}.
\newblock


\bibitem[\protect\citeauthoryear{Exter, Caskurlu, and Fernandez}{Exter
  et~al\mbox{.}}{2018}]%
        {Exter2018}
\bibfield{author}{\bibinfo{person}{Marisa Exter}, \bibinfo{person}{Secil
  Caskurlu}, {and} \bibinfo{person}{Todd Fernandez}.}
  \bibinfo{year}{2018}\natexlab{}.
\newblock \showarticletitle{{Comparing Computing Professionals’ Perceptions
  of Importance of Skills and Knowledge on the Job and Coverage in
  Undergraduate Experiences}}.
\newblock \bibinfo{journal}{\emph{ACM Transactions on Computing Education
  (TOCE)}} \bibinfo{volume}{18}, \bibinfo{number}{4} (\bibinfo{year}{2018}),
  \bibinfo{pages}{1--29}.
\newblock


\bibitem[\protect\citeauthoryear{Fakhoury, Ma, Arnaoudova, and
  Adesope}{Fakhoury et~al\mbox{.}}{2018}]%
        {Fakhoury2018}
\bibfield{author}{\bibinfo{person}{Sarah Fakhoury}, \bibinfo{person}{Yuzhan
  Ma}, \bibinfo{person}{Venera Arnaoudova}, {and} \bibinfo{person}{Olusola
  Adesope}.} \bibinfo{year}{2018}\natexlab{}.
\newblock \showarticletitle{{The Effect of Poor Source Code Lexicon and
  Readability on Developers' Cognitive Load}}. In
  \bibinfo{booktitle}{\emph{Proc.\ Int'l Conf.\ Program Comprehension (ICPC)}}.
  \bibinfo{publisher}{IEEE}, \bibinfo{pages}{286--28610}.
\newblock


\bibitem[\protect\citeauthoryear{Fakhoury, Roy, Ma, Arnaoudova, and
  Adesope}{Fakhoury et~al\mbox{.}}{2020}]%
        {Fakhoury2019}
\bibfield{author}{\bibinfo{person}{Sarah Fakhoury}, \bibinfo{person}{Devjeet
  Roy}, \bibinfo{person}{Yuzhan Ma}, \bibinfo{person}{Venera Arnaoudova}, {and}
  \bibinfo{person}{Olusola Adesope}.} \bibinfo{year}{2020}\natexlab{}.
\newblock \showarticletitle{{Measuring the Impact of Lexical and Structural
  Inconsistencies on Developers? Cognitive Load during Bug Localization}}.
\newblock \bibinfo{journal}{\emph{Empirical Softw.\ Eng.}} \bibinfo{number}{3}
  (\bibinfo{year}{2020}), \bibinfo{pages}{2140--2178}.
\newblock
Issue 25.


\bibitem[\protect\citeauthoryear{Floyd, Santander, and Weimer}{Floyd
  et~al\mbox{.}}{2017}]%
        {Floyd2017}
\bibfield{author}{\bibinfo{person}{Benjamin Floyd}, \bibinfo{person}{Tyler
  Santander}, {and} \bibinfo{person}{Westley Weimer}.}
  \bibinfo{year}{2017}\natexlab{}.
\newblock \showarticletitle{{Decoding the Representation of Code in the Brain:
  An fMRI Study of Code Review and Expertise}}. In
  \bibinfo{booktitle}{\emph{Proc.\ Int'l Conf.\ Software Engineering (ICSE)}}.
  \bibinfo{publisher}{IEEE}, \bibinfo{pages}{175--186}.
\newblock


\bibitem[\protect\citeauthoryear{Garousi, Giray, Tuzun, Catal, and
  Felderer}{Garousi et~al\mbox{.}}{2019}]%
        {Garousi2019}
\bibfield{author}{\bibinfo{person}{Vahid Garousi}, \bibinfo{person}{Gorkem
  Giray}, \bibinfo{person}{Eray Tuzun}, \bibinfo{person}{Cagatay Catal}, {and}
  \bibinfo{person}{Michael Felderer}.} \bibinfo{year}{2019}\natexlab{}.
\newblock \showarticletitle{{Closing the Gap Between Software Engineering
  Education and Industrial Needs}}.
\newblock \bibinfo{journal}{\emph{IEEE Software}} \bibinfo{volume}{37},
  \bibinfo{number}{2} (\bibinfo{year}{2019}), \bibinfo{pages}{68--77}.
\newblock


\bibitem[\protect\citeauthoryear{Ghory}{Ghory}{2007}]%
        {ghory2022}
\bibfield{author}{\bibinfo{person}{Imran Ghory}.}
  \bibinfo{year}{2007}\natexlab{}.
\newblock \bibinfo{title}{{Using FizzBuzz to Find Developers who Grok Coding}}.
\newblock
\newblock
\urldef\tempurl%
\url{https://imranontech.com/2007/01/24/using-fizzbuzz-to-find-developers-who-grok-coding/}
\showURL{%
\tempurl}


\bibitem[\protect\citeauthoryear{Gilmore}{Gilmore}{1990}]%
        {Gilmore1990}
\bibfield{author}{\bibinfo{person}{David Gilmore}.}
  \bibinfo{year}{1990}\natexlab{}.
\newblock \showarticletitle{{Expert Programming Knowledge: A Strategic
  Approach}}.
\newblock \bibinfo{journal}{\emph{Psychology of Programming}}
  (\bibinfo{year}{1990}), \bibinfo{pages}{223--234}.
\newblock


\bibitem[\protect\citeauthoryear{Gilmore and Green}{Gilmore and Green}{1988}]%
        {Gilmore1988}
\bibfield{author}{\bibinfo{person}{David Gilmore} {and} \bibinfo{person}{Thomas
  Green}.} \bibinfo{year}{1988}\natexlab{}.
\newblock \showarticletitle{{Programming Plans and Programming Expertise}}.
\newblock \bibinfo{journal}{\emph{The Quarterly Journal of Experimental
  Psychology}} \bibinfo{volume}{40}, \bibinfo{number}{3}
  (\bibinfo{year}{1988}), \bibinfo{pages}{423--442}.
\newblock


\bibitem[\protect\citeauthoryear{Groeneveld, Vennekens, and Aerts}{Groeneveld
  et~al\mbox{.}}{2021}]%
        {Groeneveld2021}
\bibfield{author}{\bibinfo{person}{Wouter Groeneveld}, \bibinfo{person}{Joost
  Vennekens}, {and} \bibinfo{person}{Kris Aerts}.}
  \bibinfo{year}{2021}\natexlab{}.
\newblock \showarticletitle{{Identifying Non-Technical Skill Gaps in Software
  Engineering Education: What Experts Expect But Students Don’t Learn}}.
\newblock \bibinfo{journal}{\emph{ACM Transactions on Computing Education
  (TOCE)}} \bibinfo{volume}{22}, \bibinfo{number}{1} (\bibinfo{year}{2021}),
  \bibinfo{pages}{1--21}.
\newblock


\bibitem[\protect\citeauthoryear{Hessels, Niehorster, Kemner, and
  Hooge}{Hessels et~al\mbox{.}}{2017}]%
        {Hessels2017}
\bibfield{author}{\bibinfo{person}{Roy Hessels}, \bibinfo{person}{Diederick
  Niehorster}, \bibinfo{person}{Chantal Kemner}, {and} \bibinfo{person}{Ignace
  Hooge}.} \bibinfo{year}{2017}\natexlab{}.
\newblock \showarticletitle{{Noise-Robust Fixation Detection in Eye Movement
  Data: Identification by Two-Means Clustering (I2MC)}}.
\newblock \bibinfo{journal}{\emph{Behavior Research Methods}}
  \bibinfo{volume}{49}, \bibinfo{number}{5} (\bibinfo{year}{2017}),
  \bibinfo{pages}{1802--1823}.
\newblock


\bibitem[\protect\citeauthoryear{Hofmeister, Siegmund, and Holt}{Hofmeister
  et~al\mbox{.}}{2019}]%
        {Hofmeister2019}
\bibfield{author}{\bibinfo{person}{Johannes Hofmeister}, \bibinfo{person}{Janet
  Siegmund}, {and} \bibinfo{person}{Daniel Holt}.}
  \bibinfo{year}{2019}\natexlab{}.
\newblock \showarticletitle{{Shorter Identifier Names Take Longer to
  Comprehend}}.
\newblock \bibinfo{journal}{\emph{Empirical Softw.\ Eng.}}
  \bibinfo{volume}{24}, \bibinfo{number}{1} (\bibinfo{year}{2019}),
  \bibinfo{pages}{417--443}.
\newblock


\bibitem[\protect\citeauthoryear{Holm, Lukander, Korpela, Sallinen, and
  M{\"u}ller}{Holm et~al\mbox{.}}{2009}]%
        {Holm2009}
\bibfield{author}{\bibinfo{person}{Anu Holm}, \bibinfo{person}{Kristian
  Lukander}, \bibinfo{person}{Jussi Korpela}, \bibinfo{person}{Mikael
  Sallinen}, {and} \bibinfo{person}{Kiti M{\"u}ller}.}
  \bibinfo{year}{2009}\natexlab{}.
\newblock \showarticletitle{{Estimating Brain Load from the EEG}}.
\newblock \bibinfo{journal}{\emph{TheScientificWorldJOURNAL}}
  \bibinfo{volume}{9} (\bibinfo{year}{2009}), \bibinfo{pages}{639--651}.
\newblock


\bibitem[\protect\citeauthoryear{Ishida and Uwano}{Ishida and Uwano}{2019}]%
        {Ishida2019EEG}
\bibfield{author}{\bibinfo{person}{Toyomi Ishida} {and}
  \bibinfo{person}{Hidetake Uwano}.} \bibinfo{year}{2019}\natexlab{}.
\newblock \showarticletitle{{Synchronized analysis of eye movement and EEG
  during program comprehension}}. In \bibinfo{booktitle}{\emph{{Proc. Int'l
  Workshop on Eye Movements in Programming (EMIP)}}}.
  \bibinfo{publisher}{IEEE}, \bibinfo{pages}{26--32}.
\newblock


\bibitem[\protect\citeauthoryear{Jasper}{Jasper}{1958}]%
        {Jasper1958}
\bibfield{author}{\bibinfo{person}{Herbert Jasper}.}
  \bibinfo{year}{1958}\natexlab{}.
\newblock \showarticletitle{{Report of the Committee on Methods of Clinical
  Examination in Electroencephalography}}.
\newblock \bibinfo{journal}{\emph{Electroencephalogr. Clin. Neurophysiol.}}
  \bibinfo{volume}{10} (\bibinfo{year}{1958}), \bibinfo{pages}{370--375}.
\newblock


\bibitem[\protect\citeauthoryear{Jbara and Feitelson}{Jbara and
  Feitelson}{2017}]%
        {Jbara2017}
\bibfield{author}{\bibinfo{person}{Ahmad Jbara} {and} \bibinfo{person}{Dror
  Feitelson}.} \bibinfo{year}{2017}\natexlab{}.
\newblock \showarticletitle{{How Programmers Read Regular Code: A Controlled
  Experiment Using Eye Tracking}}.
\newblock \bibinfo{journal}{\emph{Empirical Softw.\ Eng.}}
  \bibinfo{volume}{22}, \bibinfo{number}{3} (\bibinfo{year}{2017}),
  \bibinfo{pages}{1440--1477}.
\newblock


\bibitem[\protect\citeauthoryear{Jorgensen, Bergersen, and Liestol}{Jorgensen
  et~al\mbox{.}}{2020}]%
        {Jorgensen2020}
\bibfield{author}{\bibinfo{person}{Magne Jorgensen}, \bibinfo{person}{Gunnar
  Bergersen}, {and} \bibinfo{person}{Knut Liestol}.}
  \bibinfo{year}{2020}\natexlab{}.
\newblock \showarticletitle{{Relations Between Effort Estimates, Skill
  Indicators, and Measured Programming Skill}}.
\newblock \bibinfo{journal}{\emph{IEEE Trans.\ Softw.\ Eng.}}
  \bibinfo{volume}{47}, \bibinfo{number}{12} (\bibinfo{year}{2020}),
  \bibinfo{pages}{2892--2906}.
\newblock


\bibitem[\protect\citeauthoryear{Kartali, Jankovi{\'{c}}, Gligorijevi{\'{c}},
  Mijovi{\'{c}}, Mijovi{\'{c}}, and Leva}{Kartali et~al\mbox{.}}{2019}]%
        {Kartali2019}
\bibfield{author}{\bibinfo{person}{Aneta Kartali}, \bibinfo{person}{Milica~M.
  Jankovi{\'{c}}}, \bibinfo{person}{Ivan Gligorijevi{\'{c}}},
  \bibinfo{person}{Pavle Mijovi{\'{c}}}, \bibinfo{person}{Bogdan
  Mijovi{\'{c}}}, {and} \bibinfo{person}{Maria~Chiara Leva}.}
  \bibinfo{year}{2019}\natexlab{}.
\newblock \showarticletitle{{Real-Time Mental Workload Estimation Using EEG}}.
  In \bibinfo{booktitle}{\emph{Human Mental Workload: Models and
  Applications}}. \bibinfo{publisher}{Springer International Publishing},
  \bibinfo{pages}{20--34}.
\newblock


\bibitem[\protect\citeauthoryear{Kleinschmager and Hanenberg}{Kleinschmager and
  Hanenberg}{2011}]%
        {Kleinschmager2011}
\bibfield{author}{\bibinfo{person}{Sebastian Kleinschmager} {and}
  \bibinfo{person}{Stefan Hanenberg}.} \bibinfo{year}{2011}\natexlab{}.
\newblock \showarticletitle{{How to Rate Programming Skills in Programming
  Experiments? A Preliminary, Exploratory, Study Based on University Marks,
  Pretests, and Self-Estimation}}. In \bibinfo{booktitle}{\emph{{Proc.\
  Workshop on Evaluation and Usability of Programming Languages and Tools}}}.
  \bibinfo{publisher}{ACM}, \bibinfo{pages}{15--24}.
\newblock


\bibitem[\protect\citeauthoryear{Koenemann and Robertson}{Koenemann and
  Robertson}{1991}]%
        {Koenemann1991}
\bibfield{author}{\bibinfo{person}{J{\"{u}}rgen Koenemann} {and}
  \bibinfo{person}{Scott Robertson}.} \bibinfo{year}{1991}\natexlab{}.
\newblock \showarticletitle{{Expert Problem Solving Strategies for Program
  Comprehension}}. In \bibinfo{booktitle}{\emph{Proc.\ Conf.\ Human Factors in
  Computing Systems (CHI)}}. \bibinfo{publisher}{ACM},
  \bibinfo{pages}{125--130}.
\newblock


\bibitem[\protect\citeauthoryear{Kosti, Georgiadis, Adamos, Laskaris,
  Spinellis, and Angelis}{Kosti et~al\mbox{.}}{2018}]%
        {Kosti2018}
\bibfield{author}{\bibinfo{person}{Makrina Kosti}, \bibinfo{person}{Kostas
  Georgiadis}, \bibinfo{person}{Dimitrios Adamos}, \bibinfo{person}{Nikos
  Laskaris}, \bibinfo{person}{Diomidis Spinellis}, {and}
  \bibinfo{person}{Lefteris Angelis}.} \bibinfo{year}{2018}\natexlab{}.
\newblock \showarticletitle{{Towards an Affordable Brain Computer Interface for
  the Assessment of Programmers' Mental Workload}}.
\newblock \bibinfo{journal}{\emph{{Int.\'l J.\ Human-Computer Studies}}}
  \bibinfo{volume}{115} (\bibinfo{year}{2018}), \bibinfo{pages}{52--66}.
\newblock


\bibitem[\protect\citeauthoryear{LaToza, Garlan, Herbsleb, and Myers}{LaToza
  et~al\mbox{.}}{2007}]%
        {Latoza2007}
\bibfield{author}{\bibinfo{person}{Thomas LaToza}, \bibinfo{person}{David
  Garlan}, \bibinfo{person}{James Herbsleb}, {and} \bibinfo{person}{Brad
  Myers}.} \bibinfo{year}{2007}\natexlab{}.
\newblock \showarticletitle{{Program Comprehension as Fact Finding}}. In
  \bibinfo{booktitle}{\emph{Proc.\ Europ.\ Software Engineering
  Conf./Foundations of Software Engineering (ESEC/FSE)}}.
  \bibinfo{publisher}{ACM}, \bibinfo{pages}{361--370}.
\newblock


\bibitem[\protect\citeauthoryear{Lee, Matteson, Hooshyar, Kim, Jung, Nam, and
  Lim}{Lee et~al\mbox{.}}{2016}]%
        {Lee2016}
\bibfield{author}{\bibinfo{person}{Seolhwa Lee}, \bibinfo{person}{Andrew
  Matteson}, \bibinfo{person}{Danial Hooshyar}, \bibinfo{person}{SongHyun Kim},
  \bibinfo{person}{JaeBum Jung}, \bibinfo{person}{GiChun Nam}, {and}
  \bibinfo{person}{Heuiseok Lim}.} \bibinfo{year}{2016}\natexlab{}.
\newblock \showarticletitle{{Comparing Programming Language Comprehension
  between Novice and Expert Programmers Using EEG Analysis}}. In
  \bibinfo{booktitle}{\emph{{Proc. Int'l Conf.\ on Bioinformatics and
  Bioengineering (BIBE)}}}. \bibinfo{publisher}{IEEE},
  \bibinfo{pages}{350--355}.
\newblock


\bibitem[\protect\citeauthoryear{Li, Ko, and Zhu}{Li et~al\mbox{.}}{2015}]%
        {Li2015}
\bibfield{author}{\bibinfo{person}{Paul~Luo Li}, \bibinfo{person}{Amy Ko},
  {and} \bibinfo{person}{Jiamin Zhu}.} \bibinfo{year}{2015}\natexlab{}.
\newblock \showarticletitle{{What Makes a Great Software Engineer?}}. In
  \bibinfo{booktitle}{\emph{Proc.\ Int'l Conf.\ Software Engineering (ICSE)}},
  Vol.~\bibinfo{volume}{1}. \bibinfo{publisher}{IEEE},
  \bibinfo{pages}{700--710}.
\newblock


\bibitem[\protect\citeauthoryear{Likert}{Likert}{1932}]%
        {Likert1932}
\bibfield{author}{\bibinfo{person}{Rensis Likert}.}
  \bibinfo{year}{1932}\natexlab{}.
\newblock \showarticletitle{{A Technique for the Measurement of Attitudes}}.
\newblock \bibinfo{journal}{\emph{Archives of Psychology}}
  \bibinfo{volume}{22}, \bibinfo{number}{140} (\bibinfo{year}{1932}),
  \bibinfo{pages}{1--55}.
\newblock


\bibitem[\protect\citeauthoryear{Lin, Liao, Hu, and Wu}{Lin
  et~al\mbox{.}}{2021}]%
        {Lin2021}
\bibfield{author}{\bibinfo{person}{Yu-Tzu Lin}, \bibinfo{person}{Yi-Zhi Liao},
  \bibinfo{person}{Xiao Hu}, {and} \bibinfo{person}{Cheng-Chih Wu}.}
  \bibinfo{year}{2021}\natexlab{}.
\newblock \showarticletitle{{EEG Activities During Program Comprehension: An
  Exploration of Cognition}}.
\newblock \bibinfo{journal}{\emph{IEEE Access}}  \bibinfo{volume}{9}
  (\bibinfo{year}{2021}), \bibinfo{pages}{120407--120421}.
\newblock


\bibitem[\protect\citeauthoryear{McConnell}{McConnell}{2011}]%
        {McConnell2011}
\bibfield{author}{\bibinfo{person}{Steve McConnell}.}
  \bibinfo{year}{2011}\natexlab{}.
\newblock \showarticletitle{{What Does 10x Mean? Measuring Variations in
  Programmer Productivity.}}
\newblock In \bibinfo{booktitle}{\emph{{Making Software}}}.
  \bibinfo{publisher}{O'Reilly \& Associates, Inc.}, \bibinfo{pages}{567--574}.
\newblock


\bibitem[\protect\citeauthoryear{Mead, Gray, Hamer, James, Sorva, Clair, and
  Thomas}{Mead et~al\mbox{.}}{2006}]%
        {Mead2006}
\bibfield{author}{\bibinfo{person}{Jerry Mead}, \bibinfo{person}{Simon Gray},
  \bibinfo{person}{John Hamer}, \bibinfo{person}{Richard James},
  \bibinfo{person}{Juha Sorva}, \bibinfo{person}{Caroline Clair}, {and}
  \bibinfo{person}{Lynda Thomas}.} \bibinfo{year}{2006}\natexlab{}.
\newblock \showarticletitle{{A Cognitive Approach to Identifying Measurable
  Milestones for Programming Skill Acquisition}}.
\newblock \bibinfo{journal}{\emph{ACM SIGCSE Bulletin}} \bibinfo{volume}{38},
  \bibinfo{number}{4} (\bibinfo{year}{2006}), \bibinfo{pages}{182--194}.
\newblock


\bibitem[\protect\citeauthoryear{Medeiros, Couceiro, Castelhano, Branco,
  Duarte, Duarte, Dur{\~a}es, Madeira, Carvalho, and Teixeira}{Medeiros
  et~al\mbox{.}}{2019}]%
        {Medeiros2019EEG}
\bibfield{author}{\bibinfo{person}{Julio Medeiros}, \bibinfo{person}{Ricardo
  Couceiro}, \bibinfo{person}{Jo{\~a}o Castelhano}, \bibinfo{person}{Castelo
  Branco}, \bibinfo{person}{Gon{\c{c}}alo Duarte}, \bibinfo{person}{Catarina
  Duarte}, \bibinfo{person}{Jo{\~a}o Dur{\~a}es}, \bibinfo{person}{Henrique
  Madeira}, \bibinfo{person}{Paulo Carvalho}, {and} \bibinfo{person}{C\'{e}sar
  Teixeira}.} \bibinfo{year}{2019}\natexlab{}.
\newblock \showarticletitle{{Software Code Complexity Assessment Using EEG
  Features}}. In \bibinfo{booktitle}{\emph{Proc.\ Int'l Conf.\ Engineering in
  Medicine and Biology Society}}. \bibinfo{publisher}{IEEE},
  \bibinfo{pages}{1413--1416}.
\newblock


\bibitem[\protect\citeauthoryear{Medeiros, Couceiro, Duarte, Dur{\~a}es,
  Castelhano, Duarte, Castelo-Branco, Madeira, de~Carvalho, and
  Teixeira}{Medeiros et~al\mbox{.}}{2021}]%
        {Medeiros2021}
\bibfield{author}{\bibinfo{person}{J{\'u}lio Medeiros},
  \bibinfo{person}{Ricardo Couceiro}, \bibinfo{person}{Gon{\c{c}}alo Duarte},
  \bibinfo{person}{Jo{\~a}o Dur{\~a}es}, \bibinfo{person}{Jo{\~a}o Castelhano},
  \bibinfo{person}{Catarina Duarte}, \bibinfo{person}{Miguel Castelo-Branco},
  \bibinfo{person}{Henrique Madeira}, \bibinfo{person}{Paulo de Carvalho},
  {and} \bibinfo{person}{C{\'e}sar Teixeira}.} \bibinfo{year}{2021}\natexlab{}.
\newblock \showarticletitle{{Can EEG Be Adopted as a Neuroscience Reference for
  Assessing Software Programmers’ Cognitive Load?}}
\newblock \bibinfo{journal}{\emph{Sensors}} \bibinfo{volume}{21},
  \bibinfo{number}{7} (\bibinfo{year}{2021}), \bibinfo{pages}{2338}.
\newblock


\bibitem[\protect\citeauthoryear{Nakagawa, Kamei, Uwano, Monden, Matsumoto, and
  German}{Nakagawa et~al\mbox{.}}{2014}]%
        {Nakagawa2014}
\bibfield{author}{\bibinfo{person}{Takao Nakagawa}, \bibinfo{person}{Yasutaka
  Kamei}, \bibinfo{person}{Hidetake Uwano}, \bibinfo{person}{Akito Monden},
  \bibinfo{person}{Kenichi Matsumoto}, {and} \bibinfo{person}{Daniel German}.}
  \bibinfo{year}{2014}\natexlab{}.
\newblock \showarticletitle{{Quantifying Programmers' Mental Workload During
  Program Comprehension Based on Cerebral Blood Flow Measurement: A Controlled
  Experiment}}. In \bibinfo{booktitle}{\emph{Proc.\ Int'l Conf.\ Software
  Engineering (ICSE)}}. \bibinfo{publisher}{ACM}, \bibinfo{pages}{448--451}.
\newblock


\bibitem[\protect\citeauthoryear{Neubauer and Fink}{Neubauer and Fink}{2009}]%
        {Neubauer2009}
\bibfield{author}{\bibinfo{person}{Aljoscha Neubauer} {and}
  \bibinfo{person}{Andreas Fink}.} \bibinfo{year}{2009}\natexlab{}.
\newblock \showarticletitle{{Intelligence and Neural Efficiency}}.
\newblock \bibinfo{journal}{\emph{Neuroscience \& Biobehavioral Reviews}}
  \bibinfo{volume}{33}, \bibinfo{number}{7} (\bibinfo{year}{2009}),
  \bibinfo{pages}{1004 -- 1023}.
\newblock


\bibitem[\protect\citeauthoryear{Nivala, Hauser, Mottok, and Gruber}{Nivala
  et~al\mbox{.}}{2016}]%
        {Nivala2016}
\bibfield{author}{\bibinfo{person}{Markus Nivala}, \bibinfo{person}{Florian
  Hauser}, \bibinfo{person}{J{\"u}rgen Mottok}, {and} \bibinfo{person}{Hans
  Gruber}.} \bibinfo{year}{2016}\natexlab{}.
\newblock \showarticletitle{{Developing Visual Expertise in Software
  Engineering: An Eye Tracking Study}}. In \bibinfo{booktitle}{\emph{Global
  Engineering Education Conference (EDUCON)}}. \bibinfo{publisher}{IEEE},
  \bibinfo{pages}{613--620}.
\newblock


\bibitem[\protect\citeauthoryear{Ortin, Rodriguez-Prieto, Pascual, and
  Garcia}{Ortin et~al\mbox{.}}{2020}]%
        {Ortin2020}
\bibfield{author}{\bibinfo{person}{Francisco Ortin}, \bibinfo{person}{Oscar
  Rodriguez-Prieto}, \bibinfo{person}{Nicolas Pascual}, {and}
  \bibinfo{person}{Miguel Garcia}.} \bibinfo{year}{2020}\natexlab{}.
\newblock \showarticletitle{{Heterogeneous Tree Structure Classification to
  Label Java Programmers According to Their Expertise Level}}.
\newblock \bibinfo{journal}{\emph{Future Generation Computer Systems}}
  \bibinfo{volume}{105} (\bibinfo{year}{2020}), \bibinfo{pages}{380--394}.
\newblock


\bibitem[\protect\citeauthoryear{Peachock, Iovino, and Sharif}{Peachock
  et~al\mbox{.}}{2017}]%
        {Peachock2017}
\bibfield{author}{\bibinfo{person}{Patrick Peachock}, \bibinfo{person}{Nicholas
  Iovino}, {and} \bibinfo{person}{Bonita Sharif}.}
  \bibinfo{year}{2017}\natexlab{}.
\newblock \showarticletitle{{Investigating Eye Movements in Natural Language
  and C++ Source Code - A Replication Experiment}}. In
  \bibinfo{booktitle}{\emph{Augmented Cognition. Neurocognition and Machine
  Learning}}. \bibinfo{publisher}{Springer}, \bibinfo{pages}{206--218}.
\newblock


\bibitem[\protect\citeauthoryear{Peirce, Gray, Simpson, MacAskill,
  H{\"o}chenberger, Sogo, Kastman, and Lindel{\o}v}{Peirce
  et~al\mbox{.}}{2019}]%
        {Peirce2019Psychopy}
\bibfield{author}{\bibinfo{person}{Jonathan Peirce}, \bibinfo{person}{Jeremy
  Gray}, \bibinfo{person}{Sol Simpson}, \bibinfo{person}{Michael MacAskill},
  \bibinfo{person}{Richard H{\"o}chenberger}, \bibinfo{person}{Hiroyuki Sogo},
  \bibinfo{person}{Erik Kastman}, {and} \bibinfo{person}{Jonas Lindel{\o}v}.}
  \bibinfo{year}{2019}\natexlab{}.
\newblock \showarticletitle{{PsychoPy2: Experiments in Behavior Made Easy}}.
\newblock \bibinfo{journal}{\emph{Behavior Research Methods}}
  \bibinfo{volume}{51}, \bibinfo{number}{1} (\bibinfo{year}{2019}),
  \bibinfo{pages}{195--203}.
\newblock


\bibitem[\protect\citeauthoryear{Peitek, Apel, Parnin, Brechmann, and
  Siegmund}{Peitek et~al\mbox{.}}{2021}]%
        {Peitek2021:Complexity}
\bibfield{author}{\bibinfo{person}{Norman Peitek}, \bibinfo{person}{Sven Apel},
  \bibinfo{person}{Chris Parnin}, \bibinfo{person}{Andr\'{e} Brechmann}, {and}
  \bibinfo{person}{Janet Siegmund}.} \bibinfo{year}{2021}\natexlab{}.
\newblock \showarticletitle{{Program Comprehension and Code Complexity Metrics:
  An fMRI Study}}.
\newblock In \bibinfo{booktitle}{\emph{Proc.\ Int'l Conf.\ Software Engineering
  (ICSE)}}. \bibinfo{publisher}{ACM}, \bibinfo{pages}{524--536}.
\newblock


\bibitem[\protect\citeauthoryear{Peitek, Siegmund, and Apel}{Peitek
  et~al\mbox{.}}{2020}]%
        {Peitek2020:Linearity}
\bibfield{author}{\bibinfo{person}{Norman Peitek}, \bibinfo{person}{Janet
  Siegmund}, {and} \bibinfo{person}{Sven Apel}.}
  \bibinfo{year}{2020}\natexlab{}.
\newblock \showarticletitle{{What Drives the Reading Order of Programmers? An
  Eye Tracking Study}}.
\newblock In \bibinfo{booktitle}{\emph{Proc.\ Int'l Conf.\ Program
  Comprehension (ICPC)}}. \bibinfo{publisher}{ACM}, \bibinfo{pages}{342–353}.
\newblock


\bibitem[\protect\citeauthoryear{Peitek, Siegmund, Parnin, Apel, and
  Brechmann}{Peitek et~al\mbox{.}}{2018}]%
        {Peitek2018:Conjoint}
\bibfield{author}{\bibinfo{person}{Norman Peitek}, \bibinfo{person}{Janet
  Siegmund}, \bibinfo{person}{Chris Parnin}, \bibinfo{person}{Sven Apel}, {and}
  \bibinfo{person}{Andr\'{e} Brechmann}.} \bibinfo{year}{2018}\natexlab{}.
\newblock \showarticletitle{{Toward Conjoint Analysis of Simultaneous
  Eye-Tracking and fMRI Data for Program-Comprehension Studies}}.
\newblock In \bibinfo{booktitle}{\emph{{Proc. Int'l Workshop on Eye Movements
  in Programming (EMIP)}}}. \bibinfo{publisher}{ACM},
  \bibinfo{pages}{1:1--1:5}.
\newblock


\bibitem[\protect\citeauthoryear{Pennington}{Pennington}{1987}]%
        {Pennington1987}
\bibfield{author}{\bibinfo{person}{Nancy Pennington}.}
  \bibinfo{year}{1987}\natexlab{}.
\newblock \showarticletitle{{Stimulus Structures and Mental Representations in
  Expert Comprehension of Computer Programs}}.
\newblock \bibinfo{journal}{\emph{Cognitive Psychology}} \bibinfo{volume}{19},
  \bibinfo{number}{3} (\bibinfo{year}{1987}), \bibinfo{pages}{295--341}.
\newblock


\bibitem[\protect\citeauthoryear{Rekrut, Sharma, Schmitt, Alexandersson, and
  Kr{\"u}ger}{Rekrut et~al\mbox{.}}{2020}]%
        {Rekrut2020Decoding}
\bibfield{author}{\bibinfo{person}{Maurice Rekrut}, \bibinfo{person}{Mansi
  Sharma}, \bibinfo{person}{Matthias Schmitt}, \bibinfo{person}{Jan
  Alexandersson}, {and} \bibinfo{person}{Antonio Kr{\"u}ger}.}
  \bibinfo{year}{2020}\natexlab{}.
\newblock \showarticletitle{Decoding Semantic Categories from EEG Activity in
  Object-Based Decision Tasks}. In \bibinfo{booktitle}{\emph{Proc.\ Int'l
  Winter Conf.\ Brain-Computer Interface (BCI)}}. \bibinfo{publisher}{IEEE},
  \bibinfo{pages}{1--7}.
\newblock


\bibitem[\protect\citeauthoryear{Roehm, Tiarks, Koschke, and Maalej}{Roehm
  et~al\mbox{.}}{2012}]%
        {Roehm2012}
\bibfield{author}{\bibinfo{person}{Tobias Roehm}, \bibinfo{person}{Rebecca
  Tiarks}, \bibinfo{person}{Rainer Koschke}, {and} \bibinfo{person}{Walid
  Maalej}.} \bibinfo{year}{2012}\natexlab{}.
\newblock \showarticletitle{{How Do Professional Developers Comprehend
  Software?}}. In \bibinfo{booktitle}{\emph{Proc.\ Int'l Conf.\ Software
  Engineering (ICSE)}}. \bibinfo{publisher}{IEEE}, \bibinfo{pages}{255--265}.
\newblock


\bibitem[\protect\citeauthoryear{Shanteau}{Shanteau}{1992}]%
        {Shanteau1992Competence}
\bibfield{author}{\bibinfo{person}{James Shanteau}.}
  \bibinfo{year}{1992}\natexlab{}.
\newblock \showarticletitle{{Competence in Experts: The Role of Task
  Characteristics}}.
\newblock \bibinfo{journal}{\emph{Organizational Behavior and Human Decision
  Processes}} \bibinfo{volume}{53}, \bibinfo{number}{2} (\bibinfo{year}{1992}),
  \bibinfo{pages}{252--266}.
\newblock


\bibitem[\protect\citeauthoryear{Shanteau and Stewart}{Shanteau and
  Stewart}{1992}]%
        {Shanteau1992}
\bibfield{author}{\bibinfo{person}{James Shanteau} {and}
  \bibinfo{person}{Thomas Stewart}.} \bibinfo{year}{1992}\natexlab{}.
\newblock \showarticletitle{{Why Study Expert Decision Making? Some Historical
  Perspectives and Comments}}.
\newblock \bibinfo{journal}{\emph{Organizational Behavior and Human Decision
  Processes}} \bibinfo{volume}{53}, \bibinfo{number}{2} (\bibinfo{year}{1992}),
  \bibinfo{pages}{95 -- 106}.
\newblock


\bibitem[\protect\citeauthoryear{Sharafi, Soh, and Gu{\'e}h{\'e}neuc}{Sharafi
  et~al\mbox{.}}{2015}]%
        {Sharafi2015Systematic}
\bibfield{author}{\bibinfo{person}{Zohreh Sharafi},
  \bibinfo{person}{Z{\'e}phyrin Soh}, {and} \bibinfo{person}{Yann-Ga{\"e}l
  Gu{\'e}h{\'e}neuc}.} \bibinfo{year}{2015}\natexlab{}.
\newblock \showarticletitle{{A Systematic Literature Review on the Usage of
  Eye-Tracking in Software Engineering}}.
\newblock \bibinfo{journal}{\emph{Information and Software Technology}}
  \bibinfo{volume}{67} (\bibinfo{year}{2015}), \bibinfo{pages}{79--107}.
\newblock


\bibitem[\protect\citeauthoryear{Sharif, Falcone, and Maletic}{Sharif
  et~al\mbox{.}}{2012}]%
        {Sharif2012}
\bibfield{author}{\bibinfo{person}{Bonita Sharif}, \bibinfo{person}{Michael
  Falcone}, {and} \bibinfo{person}{Jonathan Maletic}.}
  \bibinfo{year}{2012}\natexlab{}.
\newblock \showarticletitle{{An Eye-Tracking Study on the Role of Scan Time in
  Finding Source Code Defects}}. In \bibinfo{booktitle}{\emph{{Proc.\ Symposium
  on Eye Tracking Research \& Applications (ETRA)}}}. \bibinfo{publisher}{ACM},
  \bibinfo{pages}{381--384}.
\newblock


\bibitem[\protect\citeauthoryear{Shneiderman and Mayer}{Shneiderman and
  Mayer}{1979}]%
        {Shneiderman1979}
\bibfield{author}{\bibinfo{person}{Ben Shneiderman} {and}
  \bibinfo{person}{Richard Mayer}.} \bibinfo{year}{1979}\natexlab{}.
\newblock \showarticletitle{{Syntactic/Semantic Interactions in Programmer
  Behavior: A Model and Experimental Results}}.
\newblock \bibinfo{journal}{\emph{{Int'l J.\ Parallel Programming}}}
  \bibinfo{volume}{8}, \bibinfo{number}{3} (\bibinfo{year}{1979}),
  \bibinfo{pages}{219--238}.
\newblock


\bibitem[\protect\citeauthoryear{Siegmund}{Siegmund}{2016}]%
        {Siegmund2016}
\bibfield{author}{\bibinfo{person}{Janet Siegmund}.}
  \bibinfo{year}{2016}\natexlab{}.
\newblock \showarticletitle{{Program Comprehension: Past, Present, and
  Future}}. In \bibinfo{booktitle}{\emph{Int'l Conf. Software Analysis,
  Evolution, and Reengineering (SANER)}}. \bibinfo{publisher}{IEEE},
  \bibinfo{pages}{13--20}.
\newblock


\bibitem[\protect\citeauthoryear{Siegmund, K\"astner, Apel, Parnin, Bethmann,
  Leich, Saake, and Brechmann}{Siegmund et~al\mbox{.}}{2014a}]%
        {Siegmund2014:fMRI}
\bibfield{author}{\bibinfo{person}{Janet Siegmund}, \bibinfo{person}{Christian
  K\"astner}, \bibinfo{person}{Sven Apel}, \bibinfo{person}{Chris Parnin},
  \bibinfo{person}{Anja Bethmann}, \bibinfo{person}{Thomas Leich},
  \bibinfo{person}{Gunter Saake}, {and} \bibinfo{person}{Andr\'e Brechmann}.}
  \bibinfo{year}{2014}\natexlab{a}.
\newblock \showarticletitle{{Understanding Understanding Source Code with
  Functional Magnetic Resonance Imaging}}. In \bibinfo{booktitle}{\emph{Proc.\
  Int'l Conf.\ Software Engineering (ICSE)}}. \bibinfo{publisher}{ACM},
  \bibinfo{pages}{378--389}.
\newblock


\bibitem[\protect\citeauthoryear{Siegmund, K\"{a}stner, Liebig, Apel, and
  Hanenberg}{Siegmund et~al\mbox{.}}{2014b}]%
        {Siegmund2014:Exp}
\bibfield{author}{\bibinfo{person}{Janet Siegmund}, \bibinfo{person}{Christian
  K\"{a}stner}, \bibinfo{person}{J\"{o}rg Liebig}, \bibinfo{person}{Sven Apel},
  {and} \bibinfo{person}{Stefan Hanenberg}.} \bibinfo{year}{2014}\natexlab{b}.
\newblock \showarticletitle{{Measuring and Modeling Programming Experience}}.
\newblock \bibinfo{journal}{\emph{Empirical Softw.\ Eng.}}
  \bibinfo{volume}{19}, \bibinfo{number}{5} (\bibinfo{year}{2014}),
  \bibinfo{pages}{1299--1334}.
\newblock


\bibitem[\protect\citeauthoryear{Siegmund, Peitek, Parnin, Apel, Hofmeister,
  K\"{a}stner, Begel, Bethmann, and Brechmann}{Siegmund et~al\mbox{.}}{2017}]%
        {Siegmund2017}
\bibfield{author}{\bibinfo{person}{Janet Siegmund}, \bibinfo{person}{Norman
  Peitek}, \bibinfo{person}{Chris Parnin}, \bibinfo{person}{Sven Apel},
  \bibinfo{person}{Johannes Hofmeister}, \bibinfo{person}{Christian
  K\"{a}stner}, \bibinfo{person}{Andrew Begel}, \bibinfo{person}{Anja
  Bethmann}, {and} \bibinfo{person}{Andr{\'e} Brechmann}.}
  \bibinfo{year}{2017}\natexlab{}.
\newblock \showarticletitle{{Measuring Neural Efficiency of Program
  Comprehension}}. In \bibinfo{booktitle}{\emph{Proc.\ Europ.\ Software
  Engineering Conf./Foundations of Software Engineering (ESEC/FSE)}}.
  \bibinfo{publisher}{ACM}, \bibinfo{pages}{140--150}.
\newblock


\bibitem[\protect\citeauthoryear{Siegmund and Schumann}{Siegmund and
  Schumann}{2015}]%
        {Siegmund2015Confounding}
\bibfield{author}{\bibinfo{person}{Janet Siegmund} {and} \bibinfo{person}{Jana
  Schumann}.} \bibinfo{year}{2015}\natexlab{}.
\newblock \showarticletitle{{Confounding Parameters on Program Comprehension: A
  Literature Survey}}.
\newblock \bibinfo{journal}{\emph{Empirical Softw.\ Eng.}}
  \bibinfo{volume}{20}, \bibinfo{number}{4} (\bibinfo{year}{2015}),
  \bibinfo{pages}{1159--1192}.
\newblock


\bibitem[\protect\citeauthoryear{Soloway and Ehrlich}{Soloway and
  Ehrlich}{1984}]%
        {Soloway1984}
\bibfield{author}{\bibinfo{person}{Elliot Soloway} {and} \bibinfo{person}{Kate
  Ehrlich}.} \bibinfo{year}{1984}\natexlab{}.
\newblock \showarticletitle{{Empirical Studies of Programming Knowledge}}.
\newblock \bibinfo{journal}{\emph{IEEE Trans.\ Softw.\ Eng.}}
  \bibinfo{volume}{10}, \bibinfo{number}{5} (\bibinfo{year}{1984}),
  \bibinfo{pages}{595--609}.
\newblock


\bibitem[\protect\citeauthoryear{Stanislaw, Hesketh, Kanavaros, Hesketh, and
  Robinson}{Stanislaw et~al\mbox{.}}{1994}]%
        {Stanislaw1994}
\bibfield{author}{\bibinfo{person}{Harold Stanislaw}, \bibinfo{person}{Beryl
  Hesketh}, \bibinfo{person}{Sylvia Kanavaros}, \bibinfo{person}{Tim Hesketh},
  {and} \bibinfo{person}{Ken Robinson}.} \bibinfo{year}{1994}\natexlab{}.
\newblock \showarticletitle{{A Note on the Quantification of Computer
  Programming Skill}}.
\newblock \bibinfo{journal}{\emph{International Journal of Human-Computer
  Studies}} \bibinfo{volume}{41}, \bibinfo{number}{3} (\bibinfo{year}{1994}),
  \bibinfo{pages}{351--362}.
\newblock


\bibitem[\protect\citeauthoryear{Vans, von Mayrhauser, and Somlo}{Vans
  et~al\mbox{.}}{1999}]%
        {Vans1999}
\bibfield{author}{\bibinfo{person}{Marie Vans}, \bibinfo{person}{Anneliese von
  Mayrhauser}, {and} \bibinfo{person}{Gabriel Somlo}.}
  \bibinfo{year}{1999}\natexlab{}.
\newblock \showarticletitle{{Program Understanding Behavior During Corrective
  Maintenance of Large-Scale Software}}.
\newblock \bibinfo{journal}{\emph{Int.'l Journal of Human-Computer Studies}}
  \bibinfo{volume}{51}, \bibinfo{number}{1} (\bibinfo{year}{1999}),
  \bibinfo{pages}{31--70}.
\newblock


\bibitem[\protect\citeauthoryear{Wiedenbeck}{Wiedenbeck}{1985}]%
        {Wiedenbeck1985}
\bibfield{author}{\bibinfo{person}{Susan Wiedenbeck}.}
  \bibinfo{year}{1985}\natexlab{}.
\newblock \showarticletitle{{Novice/Expert Differences in Programming Skills}}.
\newblock \bibinfo{journal}{\emph{Int.'l Journal of Man-Machine Studies}}
  \bibinfo{volume}{23}, \bibinfo{number}{4} (\bibinfo{year}{1985}),
  \bibinfo{pages}{383--390}.
\newblock


\bibitem[\protect\citeauthoryear{Wiedenbeck, Fix, and Scholtz}{Wiedenbeck
  et~al\mbox{.}}{1993}]%
        {Wiedenbeck1993}
\bibfield{author}{\bibinfo{person}{Susan Wiedenbeck}, \bibinfo{person}{Vikki
  Fix}, {and} \bibinfo{person}{Jean Scholtz}.} \bibinfo{year}{1993}\natexlab{}.
\newblock \showarticletitle{{Characteristics of the Mental Representations of
  Novice and Expert Programmers: An Empirical Study}}.
\newblock \bibinfo{journal}{\emph{Int.'l Journal of Man-Machine Studies}}
  \bibinfo{volume}{39}, \bibinfo{number}{5} (\bibinfo{year}{1993}),
  \bibinfo{pages}{793--812}.
\newblock


\bibitem[\protect\citeauthoryear{Yeh, Gopstein, Yan, and Zhuang}{Yeh
  et~al\mbox{.}}{2017}]%
        {Yeh2017}
\bibfield{author}{\bibinfo{person}{Martin Yeh}, \bibinfo{person}{Dan Gopstein},
  \bibinfo{person}{Yu Yan}, {and} \bibinfo{person}{Yanyan Zhuang}.}
  \bibinfo{year}{2017}\natexlab{}.
\newblock \showarticletitle{{Detecting and Comparing Brain Activity in Short
  Program Comprehension Using EEG}}. In \bibinfo{booktitle}{\emph{Frontiers in
  Education Conference}}. \bibinfo{publisher}{IEEE}, \bibinfo{pages}{1--5}.
\newblock


\bibitem[\protect\citeauthoryear{Zha and Wang}{Zha and Wang}{2021}]%
        {Zha2021}
\bibfield{author}{\bibinfo{person}{Fanghui Zha} {and} \bibinfo{person}{Yong
  Wang}.} \bibinfo{year}{2021}\natexlab{}.
\newblock \showarticletitle{{Correlation Analysis of Subject Competition and
  Programming Ability for Novice Programmers}}. In
  \bibinfo{booktitle}{\emph{Proc.\ Int'l Symp.\ System and Software Reliability
  (ISSSR)}}. \bibinfo{publisher}{IEEE}, \bibinfo{pages}{25--31}.
\newblock


\end{thebibliography}

\end{document}